\documentclass[twocolumn]{aastex62}%
\usepackage{graphicx}
\usepackage{enumerate}
\usepackage{lipsum}
\usepackage{mathrsfs}
\usepackage{amsmath}
\usepackage{amssymb}
\usepackage[figuresright]{rotating}
\usepackage{bm}
\usepackage{float}
\usepackage{color}

\graphicspath{{./}{figures/}}

\def\planet{Protostars and Planets VI}

\shorttitle{Numerical Simulation and Completeness Survey of Bubbles in Taurus and Perseus Molecular Cloud}
\shortauthors{Liu et al.}
\begin{document}

\title{ Numerical Simulation and Completeness Survey of Bubbles in the Taurus and Perseus Molecular Clouds }

\correspondingauthor{Mengting Liu, Di Li}
\email{liumengting@nao.cas.cn}
\email{dili@nao.cas.cn}

\author{Mengting Liu}
\affil{CAS Key Laboratory of FAST, National Astronomical Observatories, Chinese Academy of Sciences, Beijing 100101, People?s Republic of China}
\affil{University of Chinese Academy of Sciences, Beijing 100049, People?s Republic of China}

\author{Di Li}
\affil{CAS Key Laboratory of FAST, National Astronomical Observatories, Chinese Academy of Sciences, Beijing 100101, People?s Republic of China}
\affil{University of Chinese Academy of Sciences, Beijing 100049, People?s Republic of China}
\affil{NAOC-UKZN Computational Astrophysics Centre, University of KwaZulu-Natal, Durban 4000, South Africa }

\author{Marko Kr\v{c}o}
\affil{CAS Key Laboratory of FAST, National Astronomical Observatories, Chinese Academy of Sciences, Beijing 100101, People?s Republic of China}

\author{Luis C. Ho}
\affil{Kavli Institute for Astronomy and Astrophysics, Peking University, Beijing 100871, People?s Republic of China} 
\affil{Department of Astronomy, School of Physics, Peking University, Beijing 100871, People?s Republic of China}

\author{Duo Xu}
\affil{Department of Astronomy, The University of Texas at Austin, Austin, TX 78712, USA} 

\author{Huixian Li}
\affil{CAS Key Laboratory of FAST, National Astronomical Observatories, Chinese Academy of Sciences, Beijing 100101, People?s Republic of China}

%\author{Mengting Liu $^{1,2*}$, Di Li$^{1,2,3*}$, Marko Kr\v{c}o$^{1}$ , Luis C. Ho$^{4,5}$, Duo Xu$^{6}$, Huixian Li$^{1}$}

%\affiliation{
%$^1$ CAS Key Laboratory of FAST, National Astronomical Observatories, Chinese Academy of Sciences, Beijing 100101, People's Republic of China\\
%$^2$ University of Chinese Academy of Sciences, Beijing 100049, People's Republic of China\\
%$^3$ NAOC-UKZN Computational Astrophysics Centre, University of KwaZulu-Natal, Durban 4000, South Africa\\
%$^4$ Kavli Institute for Astronomy and Astrophysics, Peking University, Beijing 100871, People's Republic of China\\
%$^5$ Department of Astronomy, School of Physics, Peking University, Beijing 100871, People's Republic of China\\
%$^6$ Department of Astronomy, The University of Texas at Austin, Austin, TX 78712, USA\\
%}



\begin{abstract}

Previous studies have analyzed the energy injection into the interstellar medium due to molecular bubbles.
They found that the total kinetic energies of bubbles are comparable to, or even larger than, those of outflows but still less than the gravitational potential and turbulence energies of the hosting clouds.
We examined the possibility that previous studies underestimated the energy injection due to being unable to detect dim or incomplete bubbles.
We simulated typical molecular bubbles and inserted them into the $^{13}$CO Five College Radio Astronomical Observatory maps of the Taurus and Perseus Molecular Clouds.
We determined bubble identification completeness by applying the same procedures to both simulated and real datasets.
We proposed a detectability function for both the Taurus and Perseus molecular clouds based on a multivariate approach. 
In Taurus, bubbles with kinetic energy less than $\sim$1$\times 10^{44}$ erg are likely to be missed.
We found that the total missing kinetic energy in Taurus is less than a couple of 10$^{44}$ erg, which only accounts for around 0.2$\%$ of the total kinetic energy of identified bubbles.
In Perseus, bubbles with kinetic energy less than $\sim$2$\times 10^{44}$ erg are likely to be missed.
We found that the total missing kinetic energy in Perseus is less than $10^{45}$ erg, which only accounts for around 1$\%$ of the total kinetic energy of identified bubbles.
We thus conclude that previous manual bubble identification routines used in Taurus and Perseus can be considered to be energetically complete.
Therefore, we confirm that energy injection from dynamic structures, namely outflows and bubbles, produced by star formation feedback are sufficient to sustain turbulence at a spatial scale from $\sim$0.1 pc to $\sim$2.8 pc.

\end{abstract}



\section{Introduction} \label{sec:intro}

Stellar feedback plays a crucial role in the dynamics and energy balance of the interstellar medium (ISM; $\&$\citep{zinnecker07}. 
Feedback associated with protostars injects momentum and energy to the parent molecular cloud, altering the velocity field and density distribution of the cloud \citep{Arce01,Arce02, Arce10}, contributing to the mass loss of the surrounding dense gas \citep{Fuller02,Arce06, Arce10, Offner15}, affecting star formation efficiency \citep{Frank14, Feder15, Pavel18}, inducing changes in the chemical composition of the impacted media \citep{Bally07}, sustaining or generating turbulence (e.g., \citealp{Fukui86, Arce01, Nakamura11a, Nakamura11b, Plunkett13, Feder15, Arce11, Feder18, Li15} --- henceforth Ar11, Fe18) resisting gravitational collapse and even disrupting the surrounding gas to limit the lifetime of their parent molecular cloud \citep{Solomon81, Arce01, Hartmann01, Arce02, Duarte-Cabral12, Plunkett13}.

The primary manifestations of stellar feedback are molecular outflows and bubbles. On the Galactic scale, superbubbles are shown to be globally important, with structures spanning hundreds of parsecs, such as chimneys and large cavities \citep{Heiles79, Norman89}. 
Abundant studies of outflows have been carried out \citep{Kwan76, Snell80, Lada81, Arce10, Nakamura11a, Nakamura11b, Narayanan12, Mott17}. 
In local to main star-forming molecular cloud complexes, relatively less attention has been paid to 'local' molecular bubbles that are hollowed out by low- to intermediate-mass stars. 
Despite being expected to be less influential than superbubbles, they are substantial in number and could inject much more energy back into their natal clouds than that from outflows (Ar11, L15, Fe18).

Molecular bubbles are partially or fully enclosed three-dimensional structures whose projections on the sky resemble partial or full rings \citep{Churchwell06}. 
Young stellar objects have sufficient stellar winds to entrain and accelerate ambient gas to sculpt spherical or ring-like cavities in their surrounding molecular clouds. 
Compared to collimated outflows, bubbles affect a larger volume of ambient molecular gas. 
Compared to supernova remnants, bubbles occur around more stars and persist for a longer period of time \citep{Matzner02, Arce11}. 
In the Taurus molecular cloud complex (TMC), the kinetic energy and energy injection rate of identified bubbles are larger than those of outflows, implying a substantial input of mechanical energy into the TMC (L15).
Due to their complex morphology, bubbles are more difficult to identify \citep{Beaumont14} in comparison to the relatively clear characteristics of outflows.
In this work, we examine the completeness of bubble identification from previous studies.

The Five College Radio Astronomical Observatory (FCRAO) CO survey \citep{Goldsmith08} remains one of the largest molecular spectral-line maps of a continuous cloud complex.
Following an empirical and iterative procedure, L15 identified 37 bubbles in the TMC. Meanwhile, Ar11 detected 12 bubbles in the Perseus molecular cloud complex (PMC).
The identified Taurus bubbles inject $\sim9.2 \times 10^{46}$ erg into the surrounding ISM.
The energy injection rate was measured to be $\sim6.4 \times 10^{33}$ erg s$^{-1}$, surprisingly comparable to the turbulence dissipation rate.
In the PMC, the kinetic energy of identified bubbles is around $\sim7.6 \times 10^{46}$ erg with an energy input rate of around $\sim8.9 \times 10^{33}$ erg s$^{-1}$, which is similar to the turbulence dissipation rate. 
Small or slowly expanding bubbles may be missed.
However, large or irregular bubbles may also be missed due to confusion.
We approach this problem with semiempirical numerical simulations of artificial molecular bubbles based on the bubble model from Cazzolato $\&$ Pineault 2005.
We generated artificial spherical and partially spherical $^{13}$CO bubbles and embedded them into the real data cubes of Taurus and Perseus.
We then carried out parameter studies including average antenna temperature, number of pixels, and expansion velocity to examine the completeness of the empirical identification procedures.

In section 2, we provide a basic description of the Taurus and Perseus data sets and the bubble identification routines of previous surveys.
The bubble model is presented in section 3, including a detailed description of our method for simulating bubbles and artificial injection into the data sets. 
In section 4, we describe the bubble detection procedures and quantify detectability. 
The properties of missing bubbles and the implication for stellar feedback in Taurus and Perseus are discussed in section 5. 
In section 6, we discuss and summarize the conclusions which may be drawn from this research.

\section{Data and empirical procedures of bubble identification}

$^{13}$CO(1-0)
 is a tracer frequently used to trace molecular hydrogen and study the dynamic morphology of the ISM due to its much smaller opacities than those of $^{12}$CO(1-0).
The Taurus $^{13}$CO(1-0) and the Perseus $^{13}$CO(1-0) datasets we used are from the FCRAO CO surveys conducted between 2003 and 2005 \citep{Ridge06a, Narayanan08,  Goldsmith08, complete11}. 

The common empirical procedures used to identify bubble structures in molecular clouds are as follows: 

1. To search the regions with circular (or arc) structures brighter than the surrounding molecular gas (Ar11, L15, Fe18). 

2. To visualize the high-velocity features of these regions by plotting a Position-Velocity ($P-V$) diagram.
If there is an expanding bubble with a red-shifted part or a blue-shifted part, we can identify the $\cup$~or~$\cap$-shaped feature on the $P-V$ diagram of each region (Ar11, L15, Fe18).

3. To search for candidate sources such as pre-main-sequence stars or protostars associated with the cloud ( Ar11, L15, Fe18).

4. To compare the bubble structures in $^{13}$CO(1-0) with an infrared map to determine whether they have similar morphologies (Ar11, L15, Fe18).

The more conditions identified bubbles satisfy, the more likely they are to be real dynamic feedback driven by protostars rather than superpositions of unrelated patterns in the cloud. 
The radius and thickness of each bubble can be estimated by fitting a Gaussian to the azimuthally averaged profile of the CO integrated intensity map (Ar11, L15).

\section{Comparison of simulated bubble and identified bubble}

We simulated spherical and partially spherical $\rm ^{13}$CO(1-0) bubbles based on the model proposed by Cazzolato $\&$ Pineault (2005) for the TMC and the PMC.
This model relies on three fundamental assumptions: 

1. The density of the surrounding medium and within the bubble is different but homogeneous.

2. The bubble expansion is isotropic. 

3. All pixels on the bubble are at the same distance from Earth.

There are eight significant parameters in this model, listed in Table 1. 
Since identified bubbles in Taurus and Perseus often have partial circular or arc-like morphology, we employedcompleteness of bubble $\beta$ as another parameter. 
The detailed bubble simulation calculation is shown in the Appendix A.

\begin{deluxetable*}{lccccc}%l means left, while c means center
\tablecolumns{4}
%\rotate
\tablecaption{Artificial Bubble Parameters \tablenotemark{a}}
\tablehead{
 \colhead{Parameters} & \colhead{Definition} & \colhead{Taurus} & \colhead{Perseus}}
\startdata
$R$  & Bubble radius & [0.28 pc - 1.90 pc] &    [0.14 pc - 2.79 pc] \\
$\Delta R$& Bubble thickness & [0.03 pc, 0.44 pc] &    [0.1 pc, 0.68 pc]\\
$\rm V_{exp}$ & Bubble expansion velocity & [1.0 km $\rm s^{-1}$ - 3.3 km $\rm s^{-1}$] & [1.2 km $\rm s^{-1}$ - 6.0 km $\rm s^{-1}$]\\
$\sigma$ & Velocity dispersion  & $\sim$1.4 km $\rm s^{-1}$ & $\sim$2.0 km $\rm s^{-1}$\\
$\beta$ & The completeness of bubble & [$90^{\circ}$, $360^{\circ}$]&[$90^{\circ}$, $360^{\circ}$]\\
$n_0'$ & Bubble H$_2$ number density  & 35 $\rm cm^{-3}$& 280 $\rm cm^{-3}$\\
     &with assumed $^{13}$CO abundance of 1.43*10$^{-6}$.&\\
      &This is estimated based on intensity of &\\
      &the high-velocity line wings and used to&  \\
      &calculate the amount of gas being moved by&\\
    &bubbles (see more explanation in&\\
      &the Discussion section).&\\
$\Delta V$  &  Velocity width of a spectrometer& 0.266 km $\rm s^{-1}$& 0.066 km $\rm s^{-1}$\\
          &channel & &\\
$a_{\rm pc}$  & The bubble physical depth along& \multicolumn{2}{l}{Calculations are presented in Appendix A of Cazzolato $\&$ Pineault 2005} \\
    &       the LOS in parsecs.& &\\      
\enddata
\tablenotemark{a}{\textbf{Note}. Parameter name, definition, values for Taurus, and values for Perseus. First five parameters are measured values from L15 and Ar11, while the last four parameters are adopted values from calculation and observation property.}
\end{deluxetable*}

These artificial bubbles were embedded individually into the Taurus and Perseus $^{13}$CO(1-0) data cubes at random positions and channels.
We compared inserted artificial bubbles with real identified bubbles by morphology and high-velocity features in Figures 1-2. 
Many of the identified bubbles in Taurus are ambiguous and difficult to identify, as in the right column of Figure 1.
The identified bubbles in Perseus tend to have a more prominent and identifiable apparent characteristics.

We simulated artificial bubbles with a similar radius, thickness, $\beta$, and expansion velocity for each cloud, as shown in the left columns of Figures 1-2. 
Circular structures brighter than the surrounding molecular gas may be identified in the intensity and channel maps.
Meanwhile, the $P-V$ diagram illustrates the $\cup$- or $\cap$-shaped features. 
We found that in general artificial bubbles are somewhat easier to detect than real ones because they are not really embedded within the surrounding environment but are added on to the original images, and their morphology is necessarily independent of the surrounding environment. 
This means that our detectability function actually presents an upper limit for the bubble detectability function with the same parameters.
We found it challenging to simulate realistic artificial bubbles in part to the ambiguous morphology of identified bubbles.

\begin{figure*}%[!htb]
%\begin{figure}[H]
\centering
%\begin{minipage}[t]{0.49\linewidth} 
%\begin{subfigure}
\includegraphics[width=0.45\textwidth]{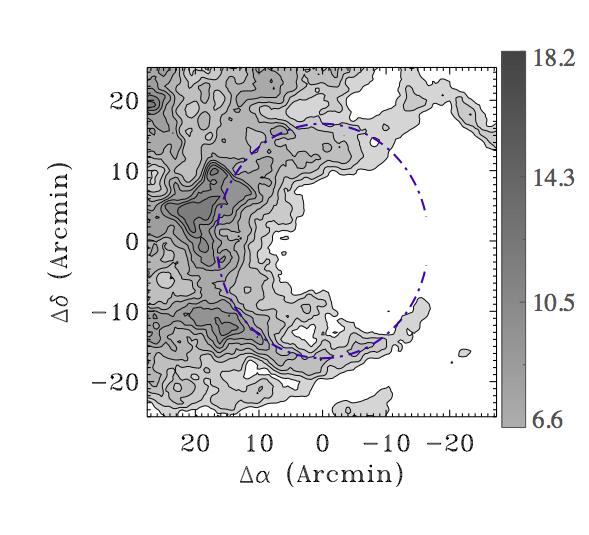}
\includegraphics[width=0.45\textwidth]{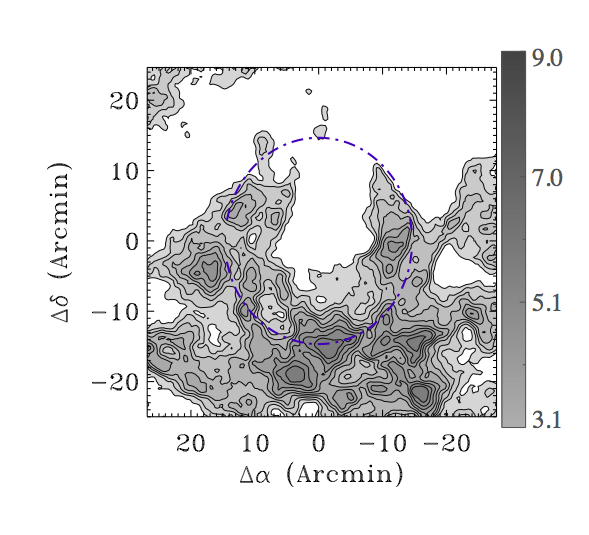}
%\end{minipage}
%\end{subfigure}
\vspace{-3mm}
\includegraphics[width=0.45\textwidth]{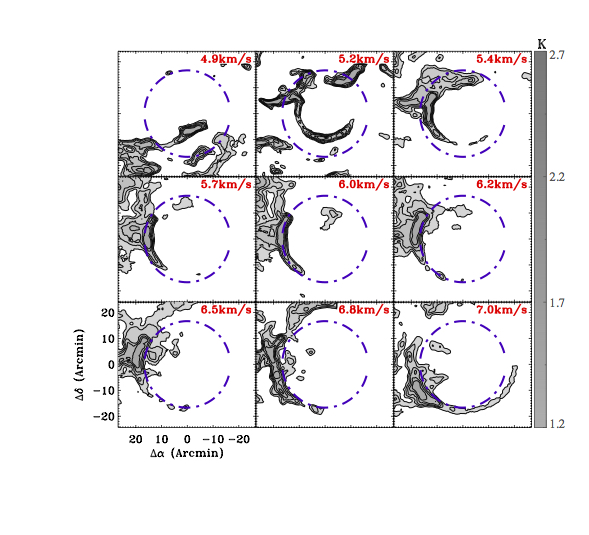}
\includegraphics[width=0.45\textwidth]{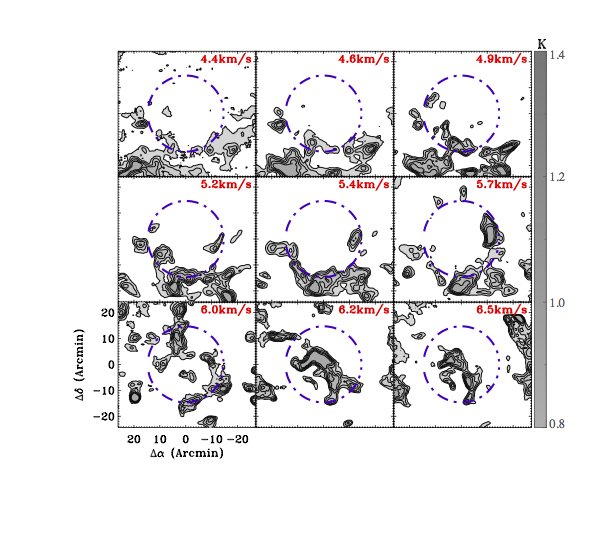}
\vspace{-3mm}
\includegraphics[width=0.35\textwidth]{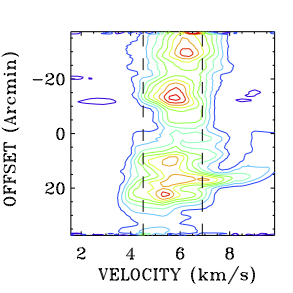}
\includegraphics[width=0.35\textwidth]{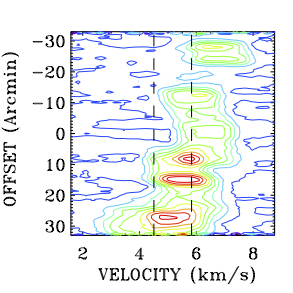}

\caption{Comparison of artificial (left column) and real (right column) bubbles in integrated intensity (top), channel maps (middle), and $P-V$ diagrams (bottom) in the TMC. 
Upper left panel: $^{13}$CO integrated intensity map of artificial bubbles inserted in the position of R.A. (J2000)=4$\rm ^{h}$38$\rm ^{m}$3$\rm ^{s}$, decl. (J2000)= 25$^{\circ}$43$'$33$''$, integrated over 4.2-6.1 km s$\rm ^{-1}$ with $R$=0.7 pc, $\beta$=260$^{\circ}$, $\Delta R$=0.2 pc, $V_{\rm exp}$=2.0 km s$^{-1}$, $n_0$=0.5 $\times$ $10^{-4}$ $cm^{-3}$, and $\sigma$=1.4 km s$^{-1}$. 
Upper right panel: $^{13}$CO integrated intensity map of identified bubble TMB$_{35}$ in position of R.A. (J2000)= 4$\rm ^{h}$46$\rm ^{m}$12$\rm ^{s}$, decl. (J2000)=25$^{\circ}$07$'$33$''$, integrated over 4.4-6.5 km s$^{-1}$ with $R$=0.62 pc and $V_{\rm exp}$=1.5 km s$^{-1}$. 
The blue dashed circles show the approximate extent of the CO emission of the bubble. Middle left panel: channel maps of the artificial bubble of $^{13}$CO emission. 
Middle right panel: channel maps of identified bubble TMB$\_{35}$ of $^{13}$CO emission. 
The blue dashed circles are the same as that in the upper panel. Lower left panel: $P-V$ diagram of artificial bubble of $^{13}$CO, through the slice shown by the angle of $110^{\circ}$. Lower right panel: $P-V$ diagram of identified bubble TMB$\_{35}$ of $^{13}$CO, through the slice shown by the angle of $110^{\circ}$. The two black dashed vertical lines show the extent of the $^{13}$CO emission associated with the bubble determined by visual inspection.} 
\end{figure*}

\begin{figure*}%[!htb]
%\begin{figure}[H]
\centering
\includegraphics[width=0.45\textwidth]{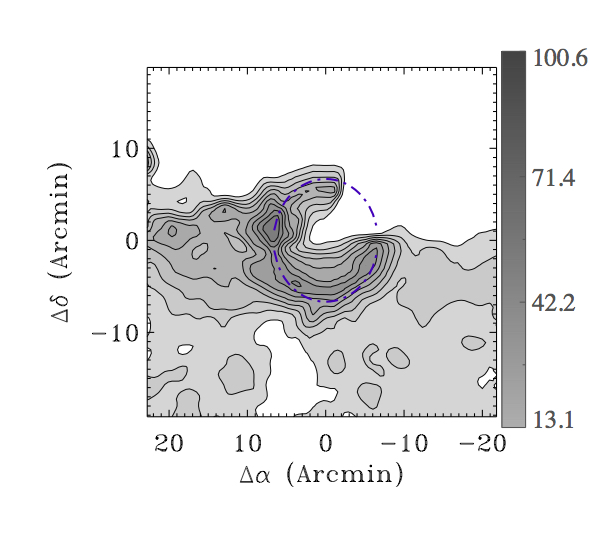}
\includegraphics[width=0.45\textwidth]{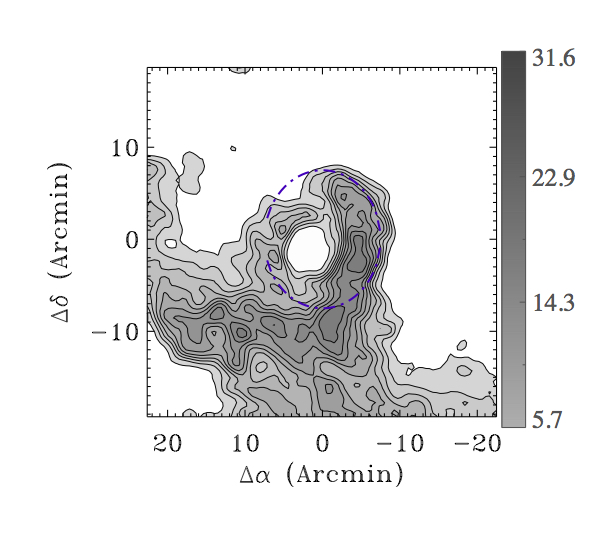}

\vspace{-3mm}

\includegraphics[width=0.45\textwidth]{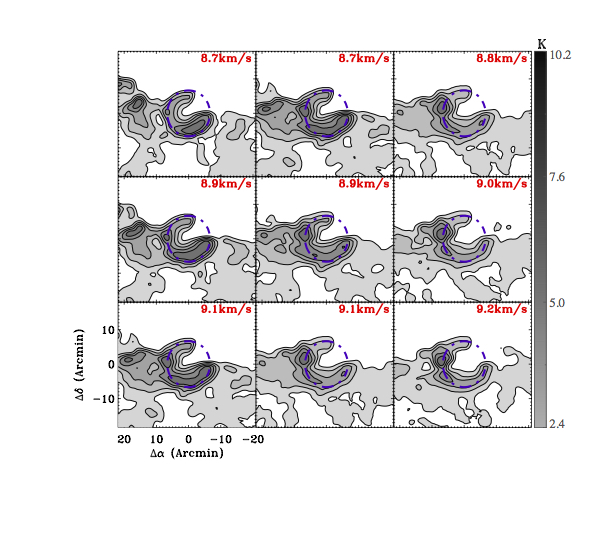}
\includegraphics[width=0.45\textwidth]{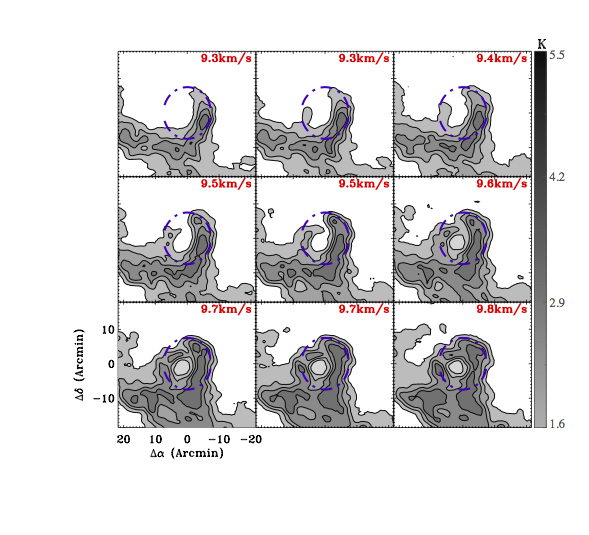}
\vspace{-3mm}

\includegraphics[width=0.35\textwidth]{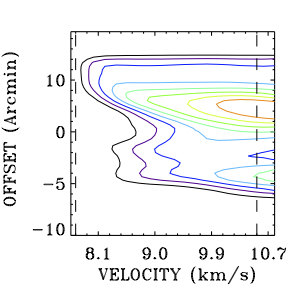}
\includegraphics[width=0.35\textwidth]{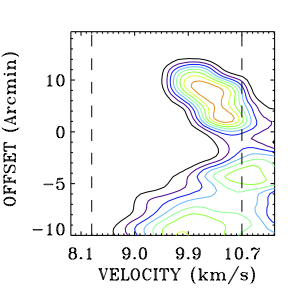}

\caption{Comparison of artificial (left column) and real (right column) bubbles in integrated intensity (top), channel maps (middle), and $P-V$ diagrams (bottom) in the PMC.
Upper left panel: $^{13}$CO integrated intensity map of artificial bubbles inserted in the position of R.A. (J2000)=3$\rm ^{h}$43$\rm ^{m}$5$\rm ^{s}$, decl. (J2000) = 32$^{\circ}$00$'$51$''$, 
integrated over 7.2-10.0 km s$^{-1}$ with $R$=0.44 pc, $\beta$=288$^{\circ}$, $\Delta R$=0.27 pc, $V_{\rm exp}$=2.5 km s$^{-1}$, $n_0$=4 $\times$ $10^{-4}$ $cm^{-3}$, and $\sigma$=2.0 km s$^{-1}$.
Upper right panel: $^{13}$CO integrated intensity map of identified bubble CPS$\_12$ in position of R.A. (J2000) = 3$\rm ^{h}$29$\rm ^{m}$26$\rm ^{s}$, delc. (J2000) = 31$^{\circ}$25$'$10$''$, integrated over 8.8-10.4 km s$^{-1}$ with $R$=0.44 pc, and $V_{\rm exp}$=2.5 km s$^{-1}$. 
The blue dashed circles show the approximate extent of the CO emission of the bubble. 
Middle left panel: channel maps of the artificial bubble of $^{13}$CO emission. 
Middle right panel: channel maps of identified bubble CPS$\_$12 of $^{13}$CO emission. 
The blue dashed circles are the same as that in the upper panel. 
Lower left panel: $P-V$ diagram of artificial bubble of $^{13}$CO, through the slice shown by the angle of 110$^{\circ}$. 
Lower right panel: $P-V$ diagram of identified bubble CPS$\_$12 of $^{13}$CO, through the slice shown by the angle of 65$^{\circ}$. 
The two black dashed vertical lines show the extent of the $^{13}$CO emission
 associated with bubble determined by visual inspection.}
\end{figure*}

\section{BUBBLE DETECTION EXPERIMENT}

Identifying the circular structure in the channel maps is the first step in bubble identification. 
Therefore, we searched for the artificial bubbles by going through the channel maps. 
When we detected a circular structure, we compared the position of detected structure with the position of the inserted artificial bubble. 
If they are consistent with each other, we consider the artificial bubble detectable. 
When we cannot detect any circular structure through channel maps, we consider the artificial bubble undetectable. 

We generated 500 artificial bubbles with different parameters for each cloud. 
Each bubble was embedded into Taurus and Perseus separately in random positions and channels one at a time.
We undertook a blind search for each inserted simulated bubble to determine the detectability function.
The artificial bubble identification routine largely follows the previous bubble manual identification procedures in section 2.

\subsection{Description of Variable Parameters}

Number of pixels $N$, average antenna temperature $T_{a}$, and expansion velocity $V{e}$ are three observable parameters which affect the bubble detection.
The number of pixels $N$ depends on the bubble radius $R$, thickness $\Delta R$, and completeness $\beta$.
The values of the parameters of each of the 500 experimental bubbles are generated randomly from the range of the maximum and minimum value of the observed bubble with uniform distribution to parameterize the bubble detectability in the whole parameter space equally.
The experimental bubbles were inserted into the Taurus and Perseus $^{13}\rm CO$ data cubes in random positions and channels one at a time.
If we cannot detect the inserted artificial bubble, the detection result of this bubble is 0, otherwise, it is 1.

\subsection{Analysis Method}
The previous bubble identification procedures are subjective and depending on 
the apparent detection threshold resulting from the bubble identification procedures in L15, Ar11, and Fe18. 
We quantified the bubble detection results of previous bubble identification procedures with the detectability function. 
Since the bubble detection results correspond to a Bernoulli distribution, which is one of the exponential family distributions, and the experimental bubble parameters can be considered as independent to each other,  we constructed the bubble detectability function using generalized linear models (GLMs; \citealp{NW72}).
GLMs are widely used regression models for dependent variables which follow an exponential family distribution, such as Gaussian distribution, Poisson distribution, and chi-squared distribution.
Derivation of the detectability function is discussed in Appendix B. 
The detectability function is in the form of
\begin{equation}
\begin{aligned}
&h(\mu_{T_{a}},\mu_{N},\mu_{V_{e}})= \\
&\dfrac{1}{1+e^{-(\alpha_{0}+\alpha_{1}*\mu_{T_{a}}+\alpha_{2}*\mu_{N}+\alpha_{3}*\mu_{V_{e}})}},\\
\end{aligned}
\end{equation}
where $\alpha_{0}$ to $\alpha_{3}$ are constants that need to be fitted,  $\mu_{T_{a}}$, $\mu_{N}$, and $\mu_{V_{e}}$ are the bubble average antenna temperature, number of pixels, and expansion velocity divided by their maximum values (see Appendix B). The maximum value of $T_{a}$, $N$, and $V_{e}$ are obtained from real observed bubbles.

\subsection{Parametric Detectability}
We performed 500 trials in both the TMC and the PMC to obtain the dataset of $N$, $T_{a}$, and $V_{e}$, and the detection result (0 or 1).
$N$, $T_{a}$, and $V_{e}$ are then divided by their maximum values. The maximum values are obtained from real observed bubbles.
We performed tenfold cross-validation on the experimental data sets to fit the constants, while testing how well the detectability function performs on the bubble identification and detectability prediction in each cloud.
In tenfold cross-validation, the data sets are randomly partitioned into 10 equal size subsets. 
One single subset is used as the validation data for testing the model, while the remaining nine subsets are used as training data to fit the model.  
We used the maximum likelihood estimator (MLE; see Appendix B) to get an estimate for each parameter from training sets and use the remaining subset for validation. We repeated this procedure 10 times. 
Our estimate for the parameters is the average over the 10 training runs, while the error is the average over the 10 validation runs.
Tenfold cross-validation is frequently used when evaluating performance of models on multiclass data \citep{G.D.13}.

\subsubsection{Results in Taurus}

We performed tenfold cross-validation on the Taurus experimental data set with random segment selection on all the data for each fold to fit the detectability function. 
We adopt the average value of 10 times the fitting results for $\alpha_{0}$ to $\alpha_{3}$, where $\alpha_{0}=-6.0$, $\alpha_{1}=112.0$, $\alpha_{2}=42.9$, and $\alpha_{3}=3.3$. 
The uncertainties are derived from the standard deviation of 10 times the fitting results for $\alpha_{0}$ to $\alpha_{3}$, which are $0.3$, $8.3$, $3.9$, and $0.5$, respectively.
The detectability function in Taurus is described as
\begin{equation}
h^T(\mu_{T_{a}},\mu_{N},\mu_{V_{e}})=\dfrac{1}{(1+e^{6.0-112.0*\mu_{T_{a}}-42.9*\mu_{N}-3.3*\mu_{V_{e}}})},
\end{equation}
where $\mu_{T_{a}}$, $\mu_{N}$, and $\mu_{V_{e}}$ are scaled experimental bubble parameters -- average antenna temperature, number of pixels, and expansion velocity -- in Taurus. 

Meanwhile, we estimated the training error from the training set and generalization error from the testing set to analyze how well $h^{T}$ performs on bubble identification and prediction in Taurus. 
The average training error is about 0.14 which is the probability that $h^{T}$ misclassifies samples in training sets. 
The average generalization error for $h^{T}$ is about 0.12 which is the probability if we draw a new set of bubble parameters and bubble detection results, $h^{T}$ misclassifies it. 
In Taurus bubble experiments, there are 330 identified experimental bubbles, which are 66$\%$ of total experiments.
As long as the detectability function can fit and predict the bubble detection result with correctness larger than 66$\%$, we can consider that the detectability function can be used for bubble identification. 
According to the average training error and generalization error, the correctness of $h^{T}$ to fit and predict the bubble detection result is 86$\%$ and 88$\%$, respectively, which indicates that $h^{T}$ can well fit and predict the bubble detection result in Taurus.

\subsubsection{Results in Perseus}

We performed tenfold cross-validation on the Perseus experimental data set with random segment selection on all the data for each fold to fit the detectability function. 
We adopt the average value of 10 times fitting results for $\alpha_{0}$ to $\alpha_{3}$, where $\alpha_{0}=-10.2$, $\alpha_{1}=121.3$, $\alpha_{2}=6.3$, and $\alpha_{3}=2.9$. 
The uncertainties are derived from the standard deviation of 10 times the fitting results for $\alpha_{0}$ to $\alpha_{3}$, which are $1.5$, $23.6$, $1.1$, and $0.2$, respectively.
The detectability function in Perseus is described as
\begin{equation}
h^{P}(\mu_{T_{a}},\mu_{N},\mu_{V_{e}})=\dfrac{1}{(1+e^{10.2-121.3*\mu_{T_{a}}-6.3* \mu_{N}-2.9*\mu_{V_{e}}})},
\end{equation}
where $\mu_{T_{a}}$, $\mu_{N}$, and $\mu_{V_{e}}$ are scaled experimental bubble parameters --- average antenna temperature, number of pixels, and expansion velocity --- in Perseus.

Meanwhile, we estimated the training error from the training set and generalization error from the testing set to analyze how well $h^{P}$ performs on bubble identification and prediction in Perseus. 
The average training error is about 0.06 which is the probability that $h^{P}$  misclassifies samples in training sets. 
The average generalization error for $h^{P}$ is about 0.08 which is the probability if we draw a new set of bubble parameters and bubble detection result, $h^{P}$ would misclassify it. 
In Perseus bubble experiments, there are 321 identified experimental bubbles, which are 64.2$\%$ of total experiments.
As long as the detectability function can fit and predict the bubble detection result with correctness larger than 64.2$\%$, we can consider that the detectability function can be used for bubble identification. 
According to the average training error and generalization error, the correctness of $h^{P}$ to fit and predict bubble detection result is 94$\%$ and 92$\%$, respectively, which indicate that $h^{P}$ can well fit and predict the bubble detection result in Perseus.

\subsection{Comparison}
The detectability functions for bubble average antenna temperature, number of pixels, and expansion velocity for each cloud are illustrated in Figure 3. 
In Taurus, the change from yellow to deep blue is more gradual than in Perseus, which is caused by the larger training error in Taurus.
We found that bubble detectability in Taurus and Perseus both strongly depend on average antenna temperature and number of pixels. 
The brightness of the bubble understandably dominates the detectability of the bubbles. In the case of a large, relatively regularly shaped bubble, the detectability roughly scaled with the number of pixels.
Weak dependence is seen to occur for expansion velocity. 
However, we still find that bubbles with larger expansion velocity are easier to detect.
In general, bubbles in Perseus are easier to detect than Taurus with the same $N$, $T_{a}$, and $V_{e}$. 
According to the behaviors of the detectability function, bubbles with low average antenna temperature, small spatial size, and
 slow expansion velocity are more likely to be missed during manual identification.
We quantify the kinetic energy of the missing bubble in the following section.

\begin{figure*}%[!htb]
%\begin{figure}[H]
\centering
\includegraphics[width=0.45\textwidth]{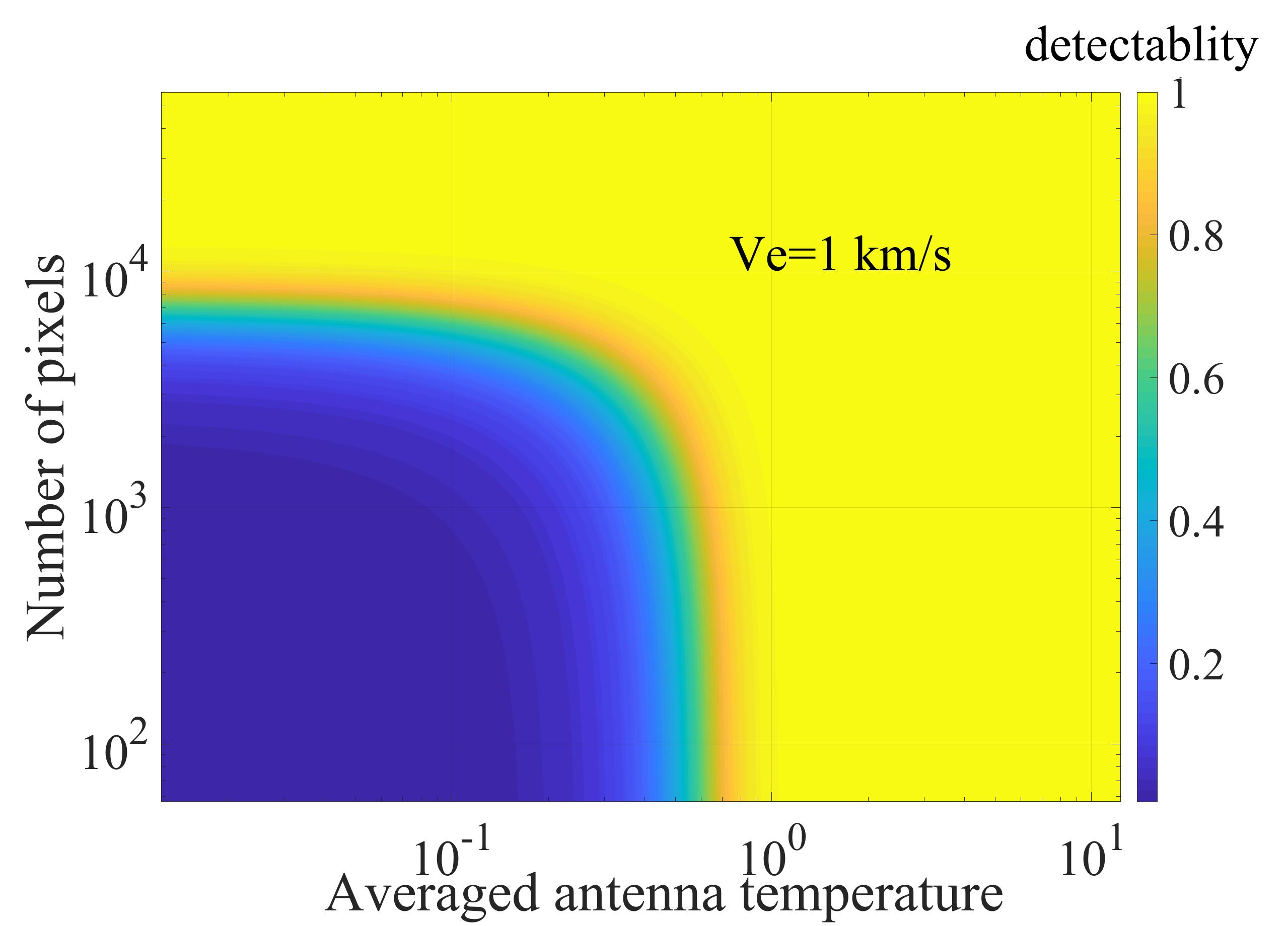}
\includegraphics[width=0.45\textwidth]{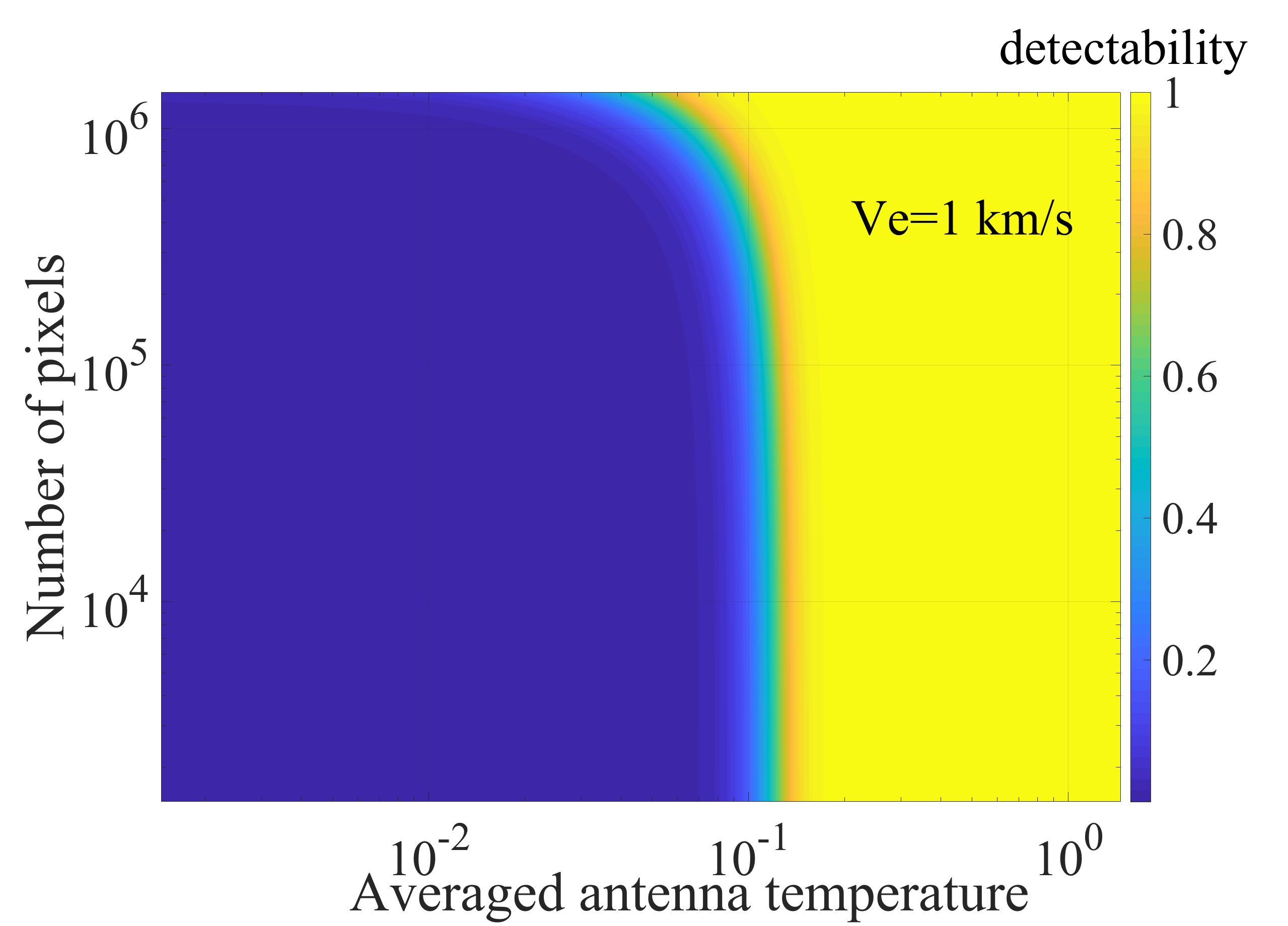}

\vspace{0.081mm}

\includegraphics[width=0.45\textwidth]{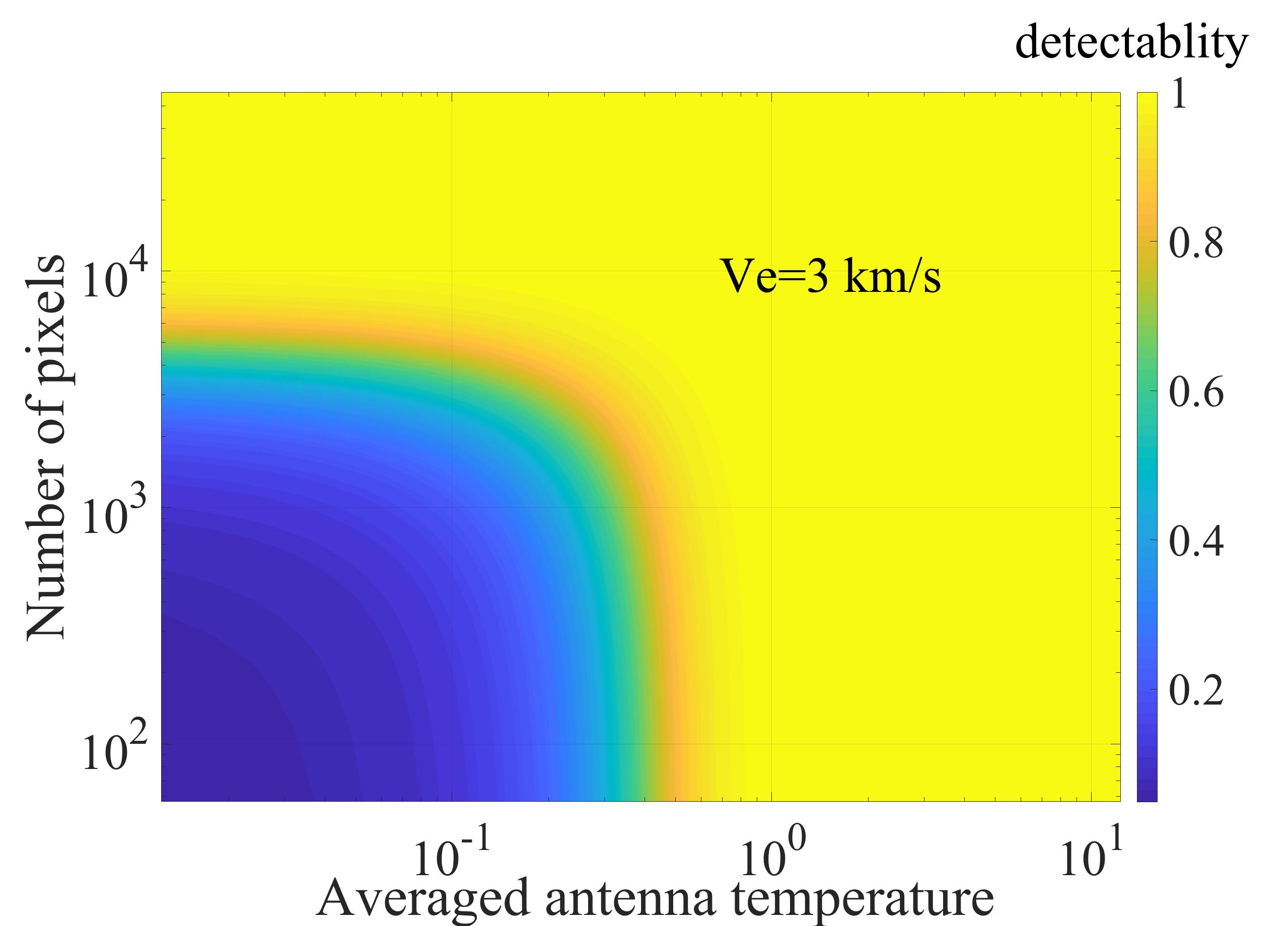}
\includegraphics[width=0.45\textwidth]{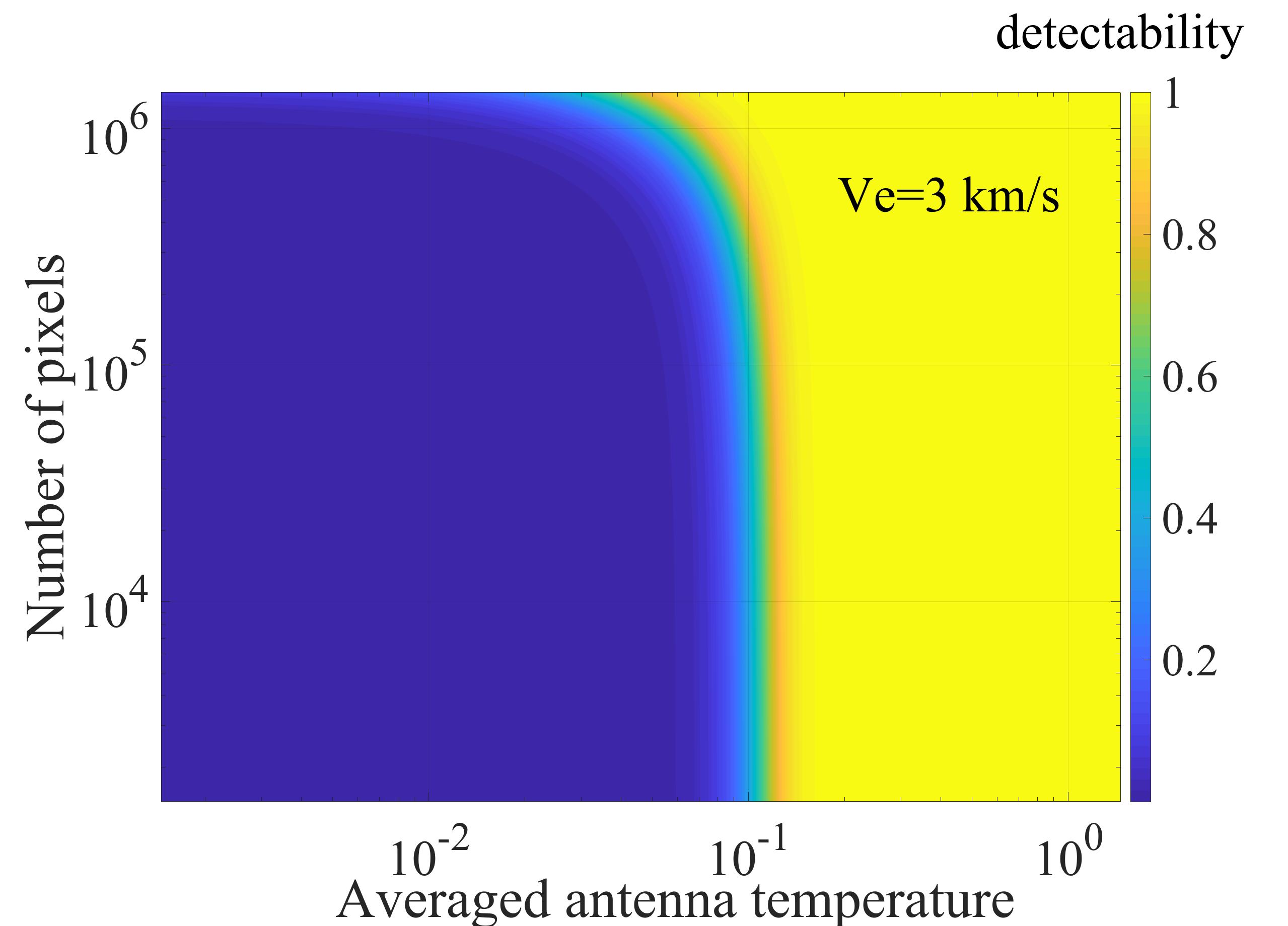}

\caption{Comparison of bubble detectability in Taurus and Perseus in different average antenna temperature, number of pixels, and expansion velocity. Left column of the diagrams displays the bubble detectability in Taurus when $V_e$ = 1 km$\rm s^{-1}$ and 3 $\rm s^{-1}$, respectively. Right column of the diagrams displays the bubble detectability in Perseus when $V_e$ = 1 km$\rm s^{-1}$ and 3 km$\rm s^{-1}$, respectively.}
\end{figure*}

\section{Kinetic energy of Missing bubbles}

Bubble kinetic energy is associated with the input of mechanical energy into the parent molecular cloud, which is the crucial parameter to evaluate the impact of bubbles on the surrounding ISM. 
We estimated the completeness of previous bubble surveys by comparing the kinetic energy of missing bubbles with identified bubbles in Taurus and Perseus. 
The identified bubbles here and afterward represent the real identified bubbles in previous bubble surveys. 

In order to estimate the kinetic energy distribution of all bubbles, including identified and missing bubbles in each molecular cloud, we assumed that the kinetic energy distribution of identified bubbles results from the kinetic energy distribution of all bubbles modified by the bubbles' detectability. Therefore, the number of the missing bubble can be estimated by the difference between the kinetic energy distribution of identified bubbles and the total number of bubbles predicted by our model.

The relation between the kinetic energy distribution of identified bubbles and all bubbles in the molecular cloud can be expressed as
\begin{equation}
\int N_{D}\cdot p(E_{k}) dE_{k} =\int N_{A}\cdot p^{tot}(E_{k})\cdot h(E_{k}) dE_{k},
\end{equation}
where $N_{D}$ is the number of real identified bubbles, $N_{A}$ is the number of all bubbles in a given cloud, $p(E_{k})$ is the probability density function (PDF) of identified bubbles with respect to kinetic energy, $p^{tot}(E_{k})$ is the PDF of all bubbles with respect to kinetic energy within the cloud, and $h(E_{k})$ denotes the detectability function with respect to kinetic energy. 

The number of missing bubbles $N^{miss}$ for kinetic energy is described as
\begin{equation}
\begin{split}
N^{miss}=\int ^{E_{k}^{max}}_{E_{k}^{min}}[N_{A}\cdot p^{tot}(E_{k})-N_{D}\cdot p(E_{k})] d E_{k},\\
=N_{D}\cdot \int ^{E_{k}^{max}}_{E_{k}^{min}}[ p(E_{k})/h(E_{k})-p(E_{k})] d E_{k}.\\
\end{split}
\end{equation}
The kinetic energy of missing bubbles $\rm E_{k}^{miss}$ is written by
\begin{equation}
\begin{split}
E_{k}^{miss}=\int ^{E_{k}^{max}}_{E_{k}^{min}}E_{k}[N_{A}\cdot p^{tot}(E_{k}) -N_{D}\cdot p(E_{k})] d E_{k},\\
=N_{D}\cdot \int ^{E_{k}^{max}}_{E_{k}^{min}}E_{k}[p(E_{k})/h(E_{k})- p(E_{k})] d E_{k}.\\
\end{split}
\end{equation}

We constructed the kinetic energy of each experimental bubble based on their number of pixels, average antenna temperature, and expansion velocity.
The detectability of each experimental bubble was evaluated by the detectability function of Equation 2 and Equation 3. 
By fitting $E_{k}$ with detectability using MLE, we estimate parameters of the kinetic energy detectability function.
The parameters of PDF of the kinetic energy of identified bubbles is estimated from the MLE. A detailed description of the fitting method is illustrated in Appendix \ref{MLE}.

\subsection{ The kinetic energy of missing bubbles in Taurus}

There are 37 identified bubbles $N_{D}$ in Taurus.
Their total kinetic energy is about 9.2$\times 10^{46}$ erg. 
The maximum and minimum kinetic energies of identified bubbles are 
%%%%%%%%%%%%%%%%%%%%%%%%%%%%%%%%%%%%%%
${{E^{max}}_{k}}^{T}$ = 4.18$\times$ $10^{46}$ erg and ${{E^{min}}_{k}}^{T}$ = 0.2$\times$ $10^{44}$ erg, respectively.
%%%%%%%%%%%%%%%%%%%%%%%
The detectability of each experimental bubble in Taurus is evaluated from Equation 2.
We plotted $E_k$ versus detectability for both experimental bubbles and identified bubbles in the right panel of Figure 4. 
Our detectability function apparently overestimates the detectability of the bubble with low kinetic energy. 
However, such overestimation does not affect energy calculation related to feedback, as these bubbles add up to negligible energies compared to bright ones. 
One possibility is that there is a large number of small, low-energy, undetectable bubbles that might contribute significant energy to the clouds. 
In order for this to be true, there would need to be at least 4600 such bubbles in the TMC and 379 in the PMC. 
We find this to be unlikely. 
By fitting $E_{k}$ with detectability using MLE, we obtain parameters of the kinetic energy detectability function by
\begin{equation}
h(E_{k})= tanh(925.5 \times E^{u}_{k}), 
\end{equation}
where ${E^{u}}_{k}$=$\dfrac{E_{k}}{{{E^{max}}_{k}}^{T}}$ refers to uniformed kinetic energy. 
We chose the tanh function because it changes from 0 to 1 gradually when kinetic energy is larger than 0, which is consistent with the detectability of all experimental bubbles from small kinetic energy to large kinetic energy. The distribution of detectability of all experimental bubbles is not a Bernoulli distribution. Therefore, we cannot use the logistic function to fit the kinetic energy detectability directly.

The bubble detectability increases as the kinetic energy increases, which makes the bubble with low kinetic energy hard to identify. The number of identified bubbles with low kinetic energy should be small. Although the bubble with high kinetic energy is easier to identify, it is unlikely that low-mass star formation in Taurus and Perseus would generate a lot of high kinetic energy bubbles. Therefore, the number of identified bubbles with high kinetic energy should be small as well.
Lognormal distribution satisfies the above conditions and can fit the real identified bubbles well.
The probability distribution of kinetic energy for identified bubbles $p(E_{k})$ can be well fitted in truncated lognormal distribution using MLE (see Appendix~\ref{sec:mlef}) by
\begin{equation}
p(E_{k}) = \frac
{
-\sqrt {2}{{\rm e}^{-\frac{1}{2}\,{\frac {1}{{\sigma}^{2}} {\left( \ln  \left(
{\frac {E_{k}}{m}} \right )  \right )} ^{2}}}}
}
{
\sqrt {\pi}\sigma\, \left( {\rm erf} \left(a_{{3}}\right )-{\rm erf}
\left(a_{{8}}\right ) \right ) E_{k}
}
\quad,
\end{equation}
where $\sigma$ is the fitting shape parameter which is 1.7$\pm0.80$ in logarithm, erf is the error function,  we defined $\mu=\ln(m)$ to be the logarithmic fitting scale parameter which is 102.8$\pm0.27$, $a_{{3}}$ and $a_{{8}}$ are compact notations which are given by
\begin{equation}
a_3 =\frac{1}{2}\,{\frac {\sqrt {2} \left( \ln  \left( {{E^{min}}_{k}}^{T} \right ) -\ln \left( m \right )  \right ) }{\sigma}}\quad , \nonumber
\end{equation}
\begin{equation}
a_8 = \frac{1}{2}\,{\frac {\sqrt {2} \left( \ln  \left( {{E^{max}}_{k}}^{T} \right ) -\ln
 \left( m \right )  \right ) }{\sigma}}
 \quad .
 \nonumber
\end{equation}
    
%%%%%%%%%%%%%%%%%%%%%%%%%%%%

%%%%%%%%%%%%%%%%%%%%%%%%%%%%
The fitting uncertainty was derived from 1000 sets of Monte Carlo experiments, each of size 37 (the number of identified bubbles in Taurus) with the kinetic energies ranging from ${{E^{min}}_{k}}^{T} $ to ${{E^{min}}_{k}}^{T} $. 
For each Monte Carlo experiment, the kinetic energy is randomly generated from the truncated lognormal distribution with $\mu=102.8$,  $\sigma=1.7$. 
We applied the same fitting algorithm to those 37 random samples to estimate $\mu$ and $\sigma$ and repeated the Monte Carlo experiment 1000 times to get sets of the estimated value of $\mu$ and $\sigma$.
We took the standard deviation of the estimated 1000 sets of $\mu$ and $\sigma$ to be the fitting uncertainty.
%%%%%%%%%%%%%%%%%%%%%%%%%%%%%
Since there are only 37 identified bubbles in Taurus, we did not plot the PDF of $E_k$ of identified bubbles, instead we plotted the cumulative distribution function (CDF) of $E_k$.
The CDF of $E_{k}$ for identified bubbles and CDF of $E_k$ of fitting distribution for Taurus are illustrated in the left panel of Figure 4. 

\begin{figure*}
%\begin{figure}[H]
%  % Requires \usepackage{graphicx}
\includegraphics[width=0.5\textwidth]{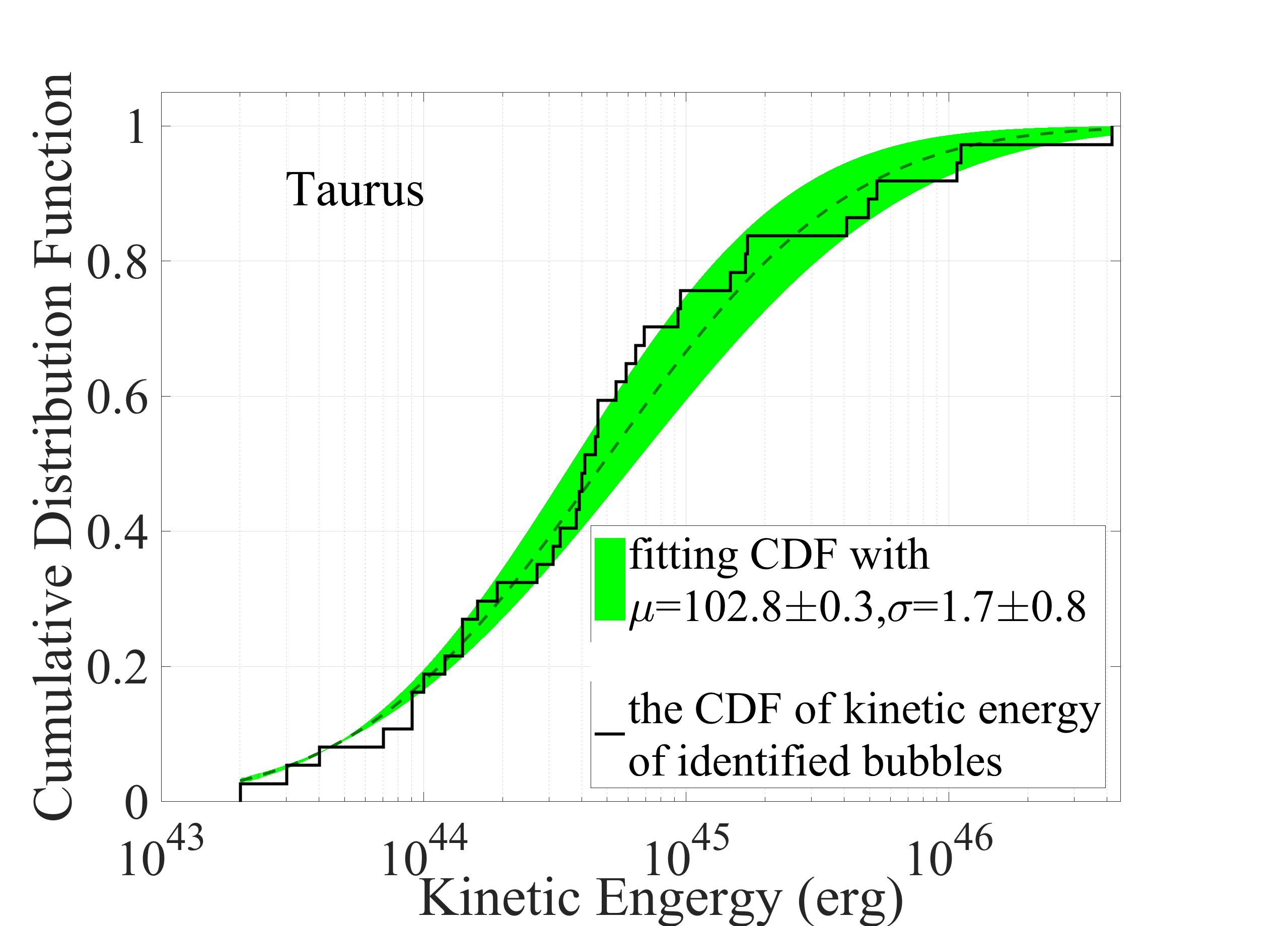}
\includegraphics[width=0.5\textwidth]{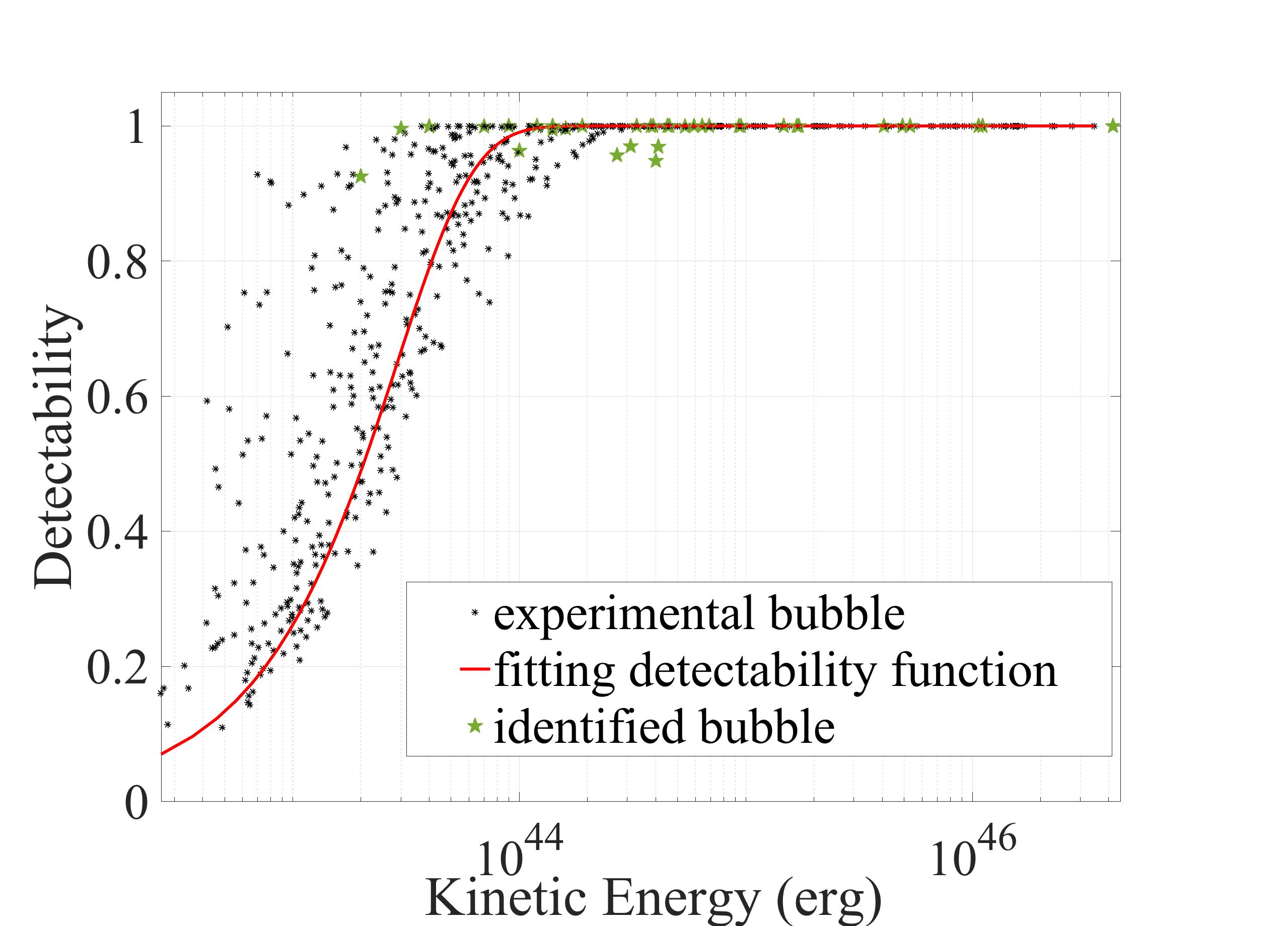}
\caption{Bubble detectability and cumulative distribution function for kinetic energy in Taurus. The left panel shows the CDF of identified bubbles for kinetic energy as black stairs. The CDF of the fitting distribution is illustrated in green, while the green area is the fitting uncertainty. The right panel shows the experimental bubble detectability as black stars and identified bubble detectability as green stars with the fitting detectability function shown in red.}
% \label{modeling-brokenbubblePV}
 \end{figure*}

By combining Equation 7 and Equation 8 with Equation 5,
the number of missing bubbles in Taurus can be derived, which ranges from 2 to 3.
This is the probable number of bubbles not identified in surveys of L15 in Taurus.
The total kinetic energy of the missing
 bubbles is estimated by combining Equation 7 and Equation 8 with Equation 6. 
However, the kinetic energy of most missing bubbles aggregates at less than $1 \times 10^{44}$ erg as illustrated in the green line of the right panel in Figure 6, which leads to the total kinetic energy of missing bubbles is ranging from 7.2$\times$10$^{43}$ erg to 8.6 $\times$10$^{43}$ erg. 
This corresponds to about 0.01$\%$ of the kinetic energy of identified bubbles. 
Therefore, the Taurus bubble survey can be considered to be energetically complete.

\subsection{Kinetic energy of missing bubbles in Perseus}

There are 12 identified bubbles $N_{D}$ in Perseus.
Their total kinetic energy is about 7.58$\times 10^{46}$ erg.
The kinetic energy of identified bubbles ranges from 2$\times 10^{44}$ erg to 1.88$\times 10^{46}$ erg.
We adopt the minimum and maximum kinetic energy of identified bubbles in Taurus (which has a wider range than in Perseus) to be the range used in estimating the missing kinetic energy in Perseus.

The detectability of each experimental bubble in Perseus is evaluated from Equation 3.
We plotted $E_k$ versus detectability for both experimental bubbles and identified bubbles in the right panel of Figure 5. 
Similar to Taurus, we obtain parameters of the kinetic energy detectability function by fitting $E_{k}$ with detectability using MLE, which can be written as
\begin{equation}
h(E_{k})=tanh(79.34\times E^{u}_{k}), 
\end{equation}
where ${E^{u}}_{k}$=$\dfrac{E_{k}}{{{E^{max}}_{k}}^{T}}$ refers to uniformed kinetic energy. 

Similarly, the probability distribution of kinetic energy for identified bubbles in Perseus can also be well fitted in truncated lognormal distribution using MLE shown in Equation 8.
The fitting shape parameter is 2.10$\pm0.43$ in logarithm. The logarithmic fitting scale parameter is 119.92$\pm0.22$.
%%%%%%%%%%%%%%%%%%%%%%%%%%%%
The fitting uncertainty was derived from 1000 sets of Monte Carlo experiments, each of size 12 (the number of identified bubbles in Perseus) with the kinetic energies ranging from ${{E^{min}}_{k}}^{P} $ to ${{E^{min}}_{k}}^{P} $. 
For each Monte Carlo experiment, the kinetic energy is randomly generated from the truncated lognormal distribution with $\mu=119.92$ and $\sigma = 2.10$. 
We applied the same fitting algorithm to those 12 random samples to estimate $\mu$ and $\sigma$ then repeated the Monte Carlo experiment 1000 times to get sets of the estimated value of $\mu$ and $\sigma$.
The fitting uncertainty is the standard deviation of the estimated 1000 sets of $\mu$ and $\sigma$.
%%%%%%%%%%%%%%%%%%%%%%%%%%%%%
There are only 12 identified bubbles in Perseus. 
Similar to Taurus, we plotted up the CDF of $E_k$ for identified bubbles and CDF of $E_k$ of the fitting distribution for Perseus, which are illustrated in the left panel of Figure 5.
%  pdf*dEk integration =1 

\begin{figure*}
%\begin{figure}[H]
%  % Requires \usepackage{graphicx}
\includegraphics[width=0.5\textwidth]{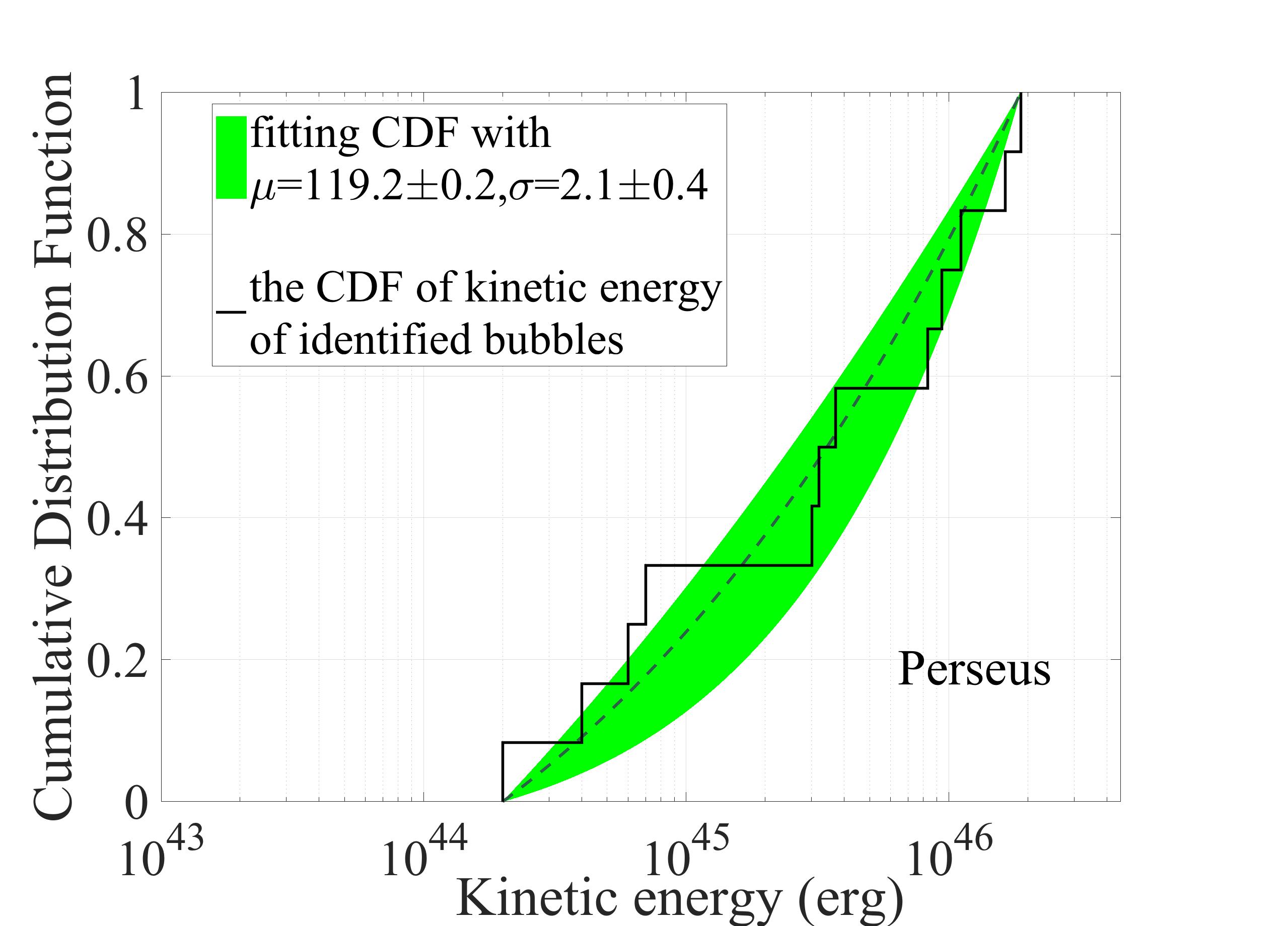}
\includegraphics[width=0.5\textwidth]{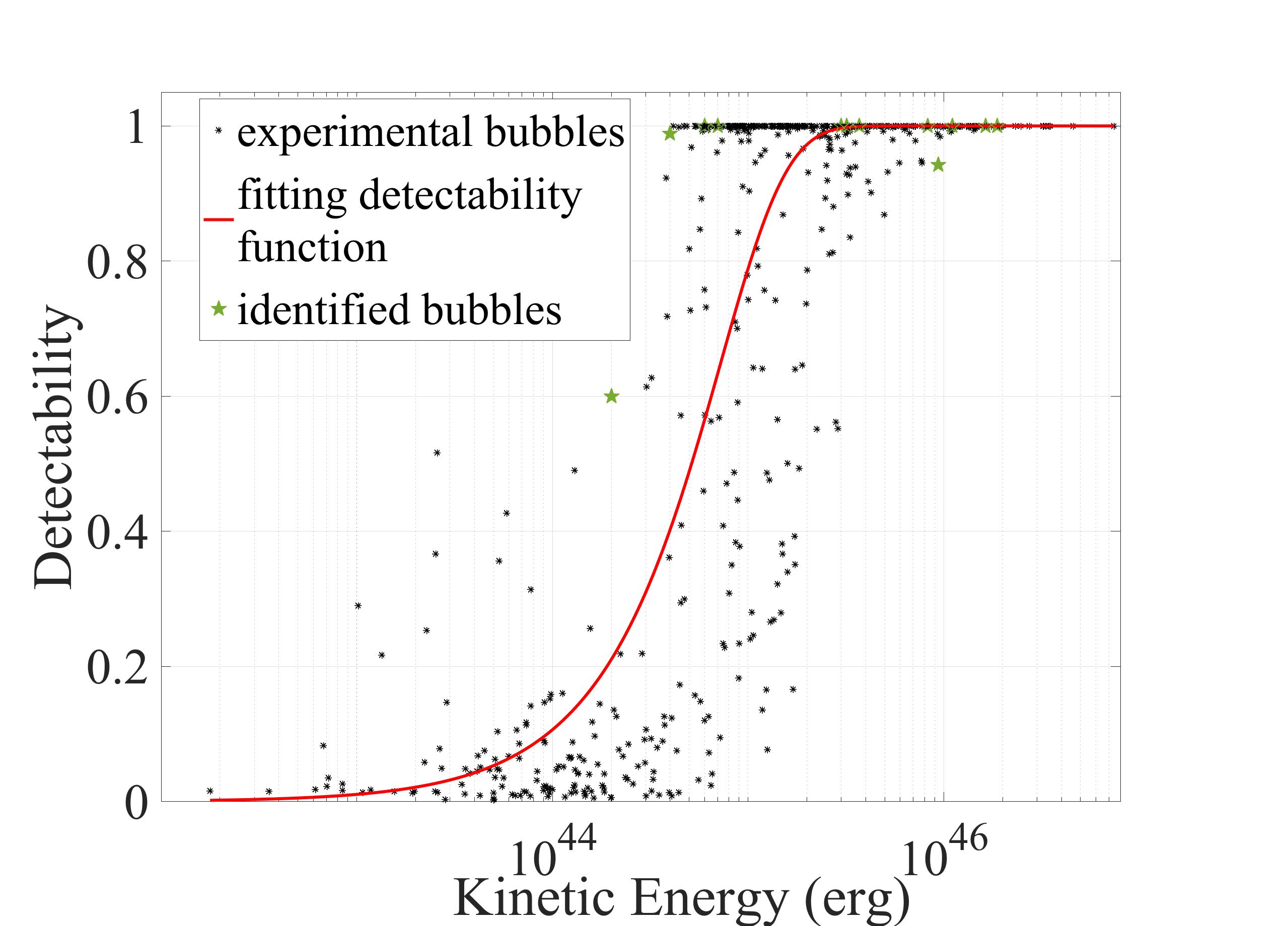}
\caption{Bubble detectability and cumulative distribution function for kinetic energy in Perseus. The left panel shows the CDF of identified bubbles for kinetic energy in black stairs. The CDF of the fitting distribution is illustrated in green, while the green area is the fitting uncertainty. The right panel shows the experimental bubble detectability as black stars and identified bubble detectability as green stars with the fitting detectability function in red.}
% \label{modeling-brokenbubblePV}
 \end{figure*}

By combining Equation 8 and Equation 9 with Equation 5, the number of missing bubbles in Perseus can be derived which is ranging from 3 to 14. 
This is the probable number of bubbles not identified in the Perseus survey by Ar11.
The total kinetic energy of the missing bubbles is estimated by combining Equation 9 and Equation 8 with Equation 6. 
However, the kinetic energy of most missing bubbles aggregates at less than $2 \times 10^{44}$ erg as illustrated in the green line of the left panel of Figure 6, which leads to the total kinetic energy of missing bubbles is ranging from 1.4$\times$10$^{44}$ erg to 8.0$\times$10$^{44}$ erg. 
This corresponds to about 1$\%$ of the kinetic energy of identified bubbles. 
Therefore, the Perseus bubble survey can be considered to be energetically complete.

\begin{figure*}
%\begin{figure}[H]
%  % Requires \usepackage{graphicx}
\includegraphics[width=0.5\textwidth]{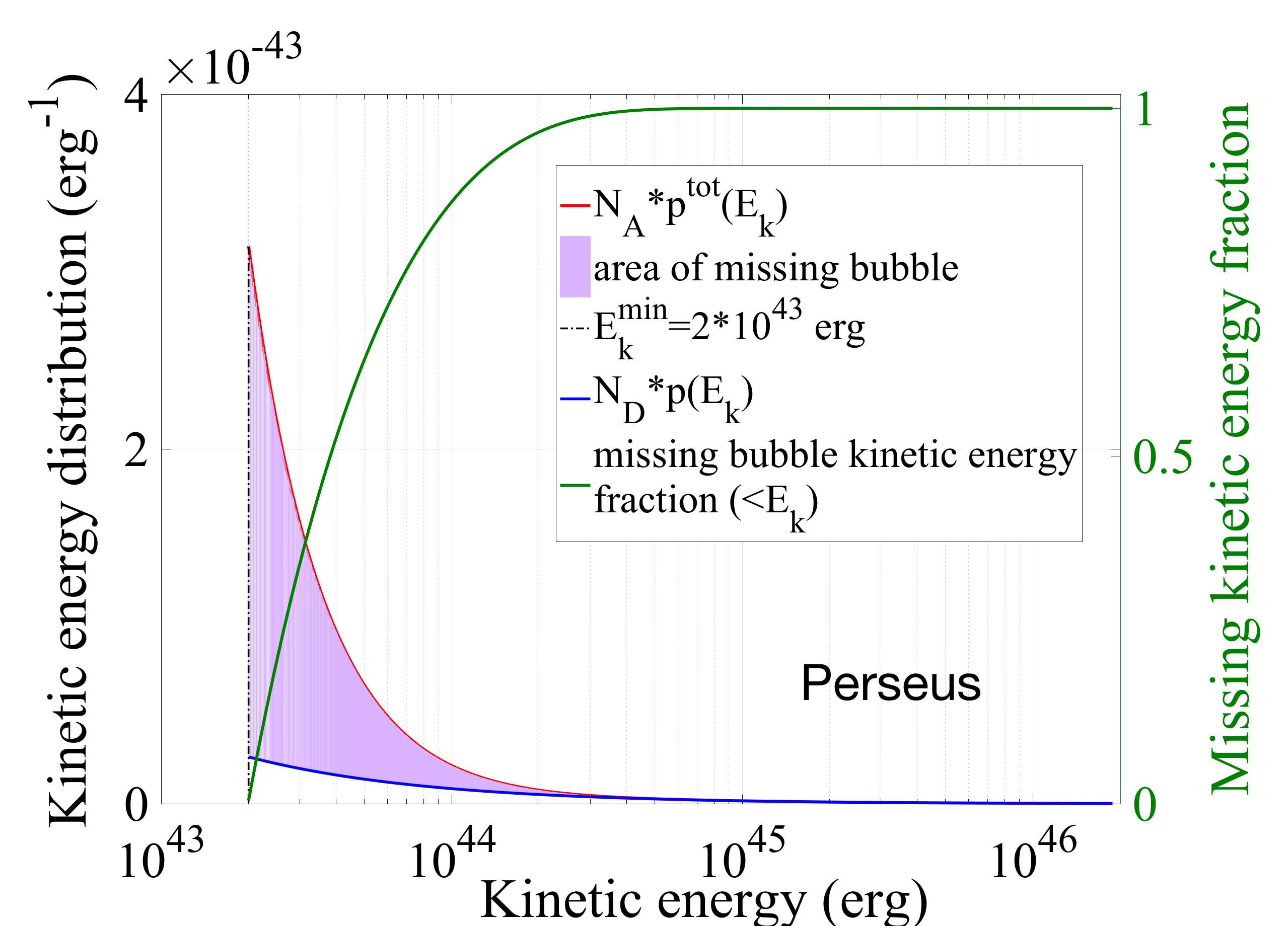}
\includegraphics[width=0.5\textwidth]{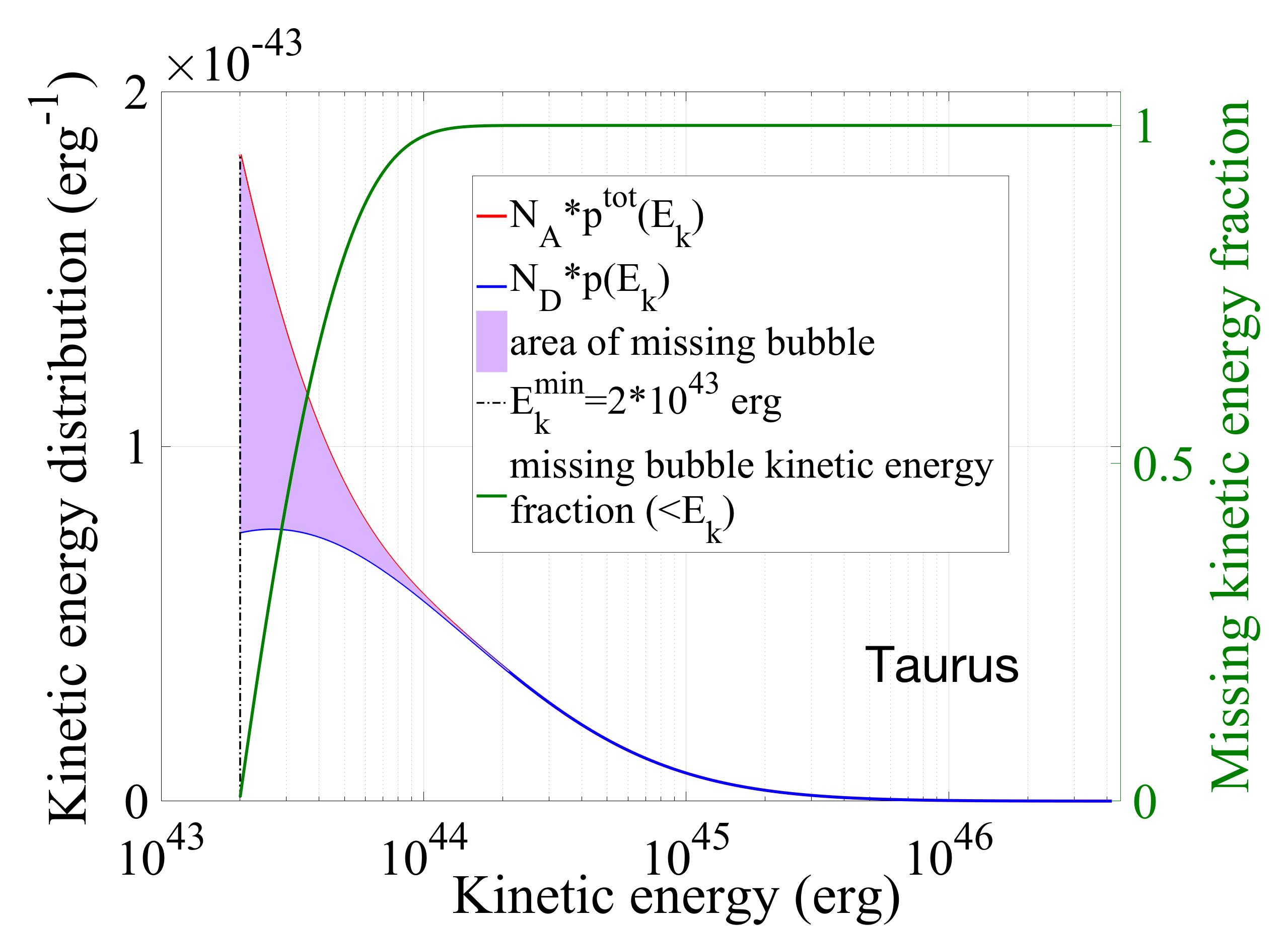}
\caption{Comparison of the distribution and missing bubble kinetic energy fraction. The left panel shows the distribution of bubbles for kinetic energy in Perseus; the red curve presents the distribution of all bubbles in Perseus for kinetic energy and the blue curve presents the distribution of identified bubbles in Perseus for kinetic energy. The purple area illustrates the missing bubble. The green curve shows the missing bubble kinetic energy fraction, which represents the percentage of missing kinetic energy lower than $E_{k}$ to the total missing kinetic energy. The right panel shows the distribution of bubbles for kinetic energy in Taurus. Same as the left panel.}
% \label{modeling-brokenbubblePV}
 \end{figure*}

 \section{Discussion and CONCLUSIONS}
Stellar feedback has a significant impact on the surrounding gas. 
The total kinetic energy of bubbles detected in Taurus and Perseus are $\sim3.9\times 10^{46}$ erg and $\sim7.6\times 10^{46}$ erg, respectively.
The gravitational potential energy and turbulence energy ($E_{G}$, $E_{\rm tur}$) of Taurus and Perseus are about
$1.5\times 10^{48}$ erg, $3.2\times 10^{47}$ erg, and $6\times 10^{47}$, $1.6 \times 10^{47}$ erg, respectively.
The energy contained in bubbles are orders of magnitude smaller than those of either gravity or turbulence, which are expected to be equal in a virialized supersonic cloud.  
The maximum energy injection (into the natal cloud) rate of molecular bubbles can be estimated from their momentum. 
They are $\sim6.4\times10^{33}$ erg $\rm s^{-1}$, and $\sim2\times10^{33}$ erg $\rm s^{-1}$ for bubbles in Taurus and Perseus, respectively. 
Following the methods laid out in Ar11 and L15, we estimated the turbulence dissipation rate of Taurus and Perseus to be $\sim3.1\times10^{33}$ erg $\rm s^{-1}$ and $\sim1\times10^{33}$ erg $\rm s^{-1}$, respectively.  
The observed injection rates are thus similar to or even slightly larger than the dissipation rates. 

Typically, when discussing molecular clouds we refer to total proton volume densities of 10$^{3}$ cm$^{-3}$ for regions with sufficient volume density to produce $^{13}$CO emission. 
This is typical for regions with A$_{v}$$\geq$1. 
However, the volumes of gas that are affected by bubbles are large and not necessarily located in the dense regions. 
In fact, they almost necessarily include large volumes of very diffuse gas. 
The gas that is dynamically affected ('moved') by bubble expansion and can be identified as such has to exhibit a velocity offset (Figure 7). 
Such 'high' velocity gas has much lower density.
Due to this, we chose to use our estimates of the average H$_2$ volume densities within Taurus and Perseus (35 cm$^{-3}$ and 280 cm$^{-3}$ respectively) when estimating the energy injected into the cloud by each bubble. 
As a comparison, if the main cloud density was assumed to be 3.5 $\times$ 10$^3$ cm$^{-3}$, the energy injected into the cloud by star formation feedback in Taurus and Perseus would be $\sim3.9\times 10^{48}$ erg and $\sim9.1\times 10^{47}$ erg, respectively, which could disperse the whole cloud. Similarly, if we use such a high volume density for the undisturbed gas in our simulations, then the resulting bubbles are much brighter than any that are actually observed.

\begin{figure}
\includegraphics[width=0.5\textwidth]{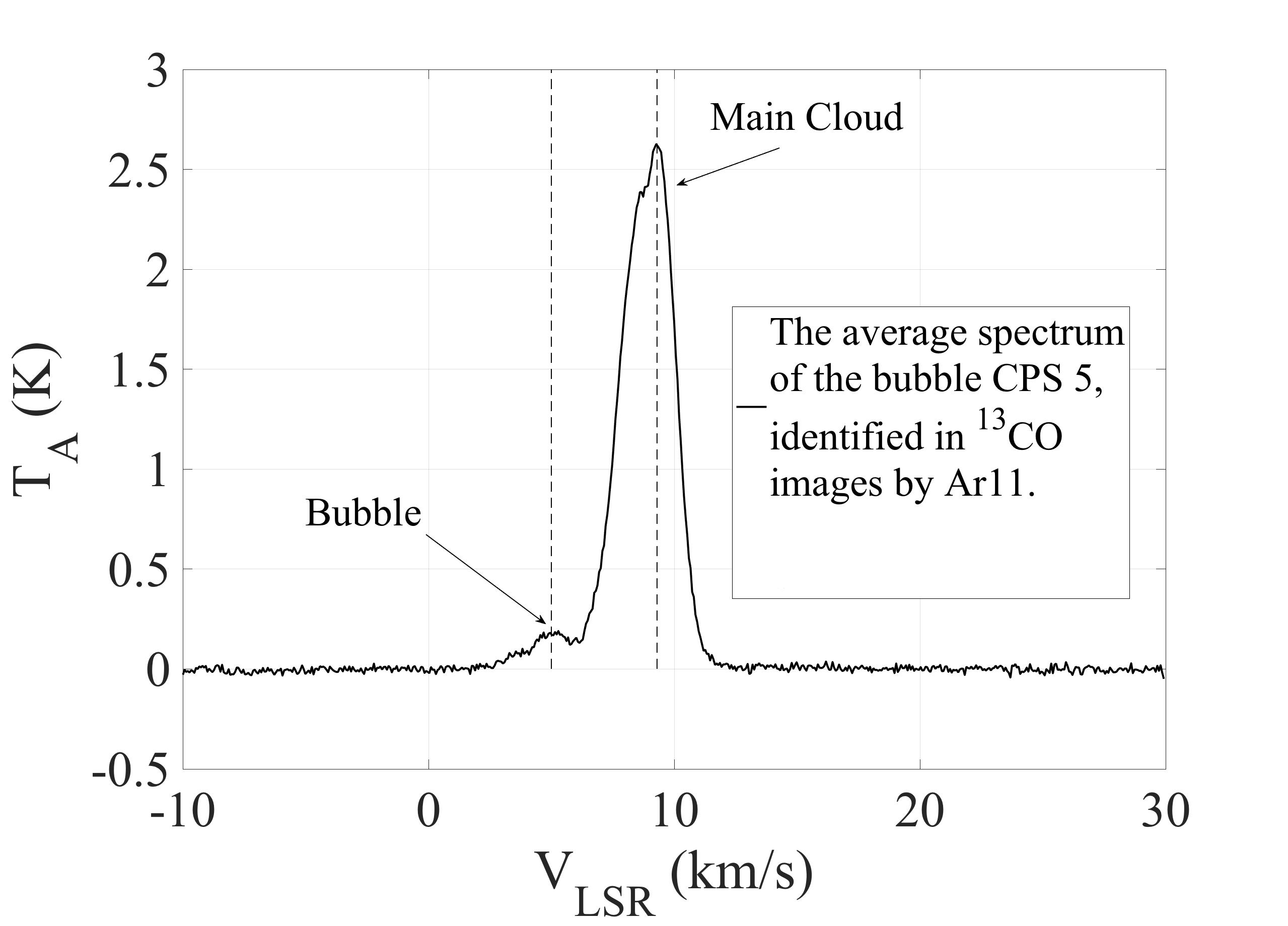}
\caption{Average spectrum of the bubble CSP 5, identified in $\rm^{13}CO$ by Ar11. The bubble was identified between 2.0 and 6.0 km $\rm s^{-1}$. The dashed lines present the bubble velocity offset from the emission of the main cloud.}
 \end{figure}

We simulated bubbles for the Taurus and Perseus molecular cloud. 
We randomly inserted artificial bubbles into the Taurus and Perseus $\rm^{13}CO$ data cubes from the 13.7 m FCRAO telescope. 
With changing average antenna temperature, the number of pixels, and expansion velocity of artificial bubbles,
we parameterized bubble detections according to the bubble detectability functions to evaluate how the detectability is affected by the parameters for two molecular clouds. 
According to the properties of identified bubbles in Taurus and Perseus and the detectability functions, we estimated the energetic completeness of previously stellar feedback studies. 
Our conclusions regarding the completeness of bubble identification
 in Taurus and Perseus and their properties are as follows:

1. The detectability of bubbles can be described as a logistic function, which is derived from GLMs. 
The distribution of bubble detection results is a Bernoulli distribution. Therefore we adopted GLMs to fit the bubble detectability functions.
In the TMC, the detectability function can be described as 

$\rm h^{T}(\mu_{T_{a}},\mu_{N},\mu_{V_{e}})=\dfrac{1}{(1+e^{6.0-112.0*\mu_{T_{a}}-42.9*\mu_{N}-3.3*\mu_{V_{e}}})}$, 
while in the PMC as 

$\rm h^{P}(\mu_{T_{a}},\mu_{N},\mu_{V_{e}})=\dfrac{1}{(1+e^{10.2-121.3*\mu_{T_{a}}-6.3* \mu_{N}-2.9*\mu_{V_{e}}})}$.
We then fitted the bubble kinetic energy distributions in Taurus and Perseus with truncated lognormal distribution.

2. The number of missing bubbles in Taurus is less than 8$\%$ of the number of identified bubbles. 
The number of missing bubbles in Perseus is about 25$\%$-125$\%$ of the number of identified bubbles.

3. We used bubble kinetic energy distributions and the bubble detectability functions to estimate the total kinetic energy of missing bubbles, which suggests that although the numbers of missing bubbles are large, their kinetic energies are relatively small (usually less than 2$\times$10$^{44}$ erg).
The total kinetic energies of missing bubbles in Taurus and Perseus during manual identification range from 7.2$\times$10$^{43}$ erg to 8.6 $\times$10$^{43}$ erg and 1.4$\times$10$^{44}$ erg to 8.0$\times$10$^{44}$ erg, respectively. 
Such potential incompleteness only accounts for $\sim0.2\%$ and $\sim1\%$ of the total kinetic energy of identified bubbles in Taurus and Perseus, respectively.

4. The empirical surveys in L15 and Ar11 for identifying bubble structures in Taurus and Perseus can be considered as energetically complete.

5. The total energy of bubbles in a cloud is orders of magnitude smaller than those
of either turbulence or gravity. The bubbles cannot generate the observed turbulence or
disperse the cloud. The observed energy injection rate from bubbles, now considered complete,
is similar to the turbulence dissipation rate. We conclude that, even in low-mass star-forming regions,
the feedback from star formation is sufficient to sustain turbulence at ranges from $\sim$0.1 pc to $\sim$2.8 pc scales.

\acknowledgments
This work is supported by the National Natural Science Foundation of China grant No.\ 11988101, No.\ 11725313, No.\ 11721303, the International Partnership Program of Chinese Academy of Sciences grant No.\ 114A11KYSB20160008, and the National Key R\&D Program of China No.\ 2016YFA0400702. 

\appendix

\section{Bubble simulation description}
\label{sec:bubblesi}
The $^{13}$CO column densities may be described by
\begin{equation}
\ N_{^{13}\rm CO} = 1.18\times 10^{15}T_{\rm b}\Delta V,
\label{equ:COColumndensity1} 
\end{equation}
where $\rm T_{b}$ is the brightness temperature of each pixel, $N_{^{13} \rm CO}$ is the column density, and $\Delta V$ represents the velocity width of a spectrometer channel which is listed in Table 1. 
This relation is derived under several assumptions. We assumed that the excitation temperature $T_{ex}$ of $^{13} \rm CO$ is 25 K, the background temperature is 2.7 K, the $^{13}\rm CO$ emission from the bubble is generally optically thin ($\tau (^{13}\rm CO) \ll 1$) and in local thermal equilibrium. 
$N_{^{13}\rm CO}$ can also be expressed as 
\begin{equation}
\ N_{^{13}\rm CO}= 3.0857 \times 10^{18} n_{\rm s} a_{\rm pc}cm^{-2}, 
\end{equation}
where $a_{\rm pc}$ describes the bubble physical depth along the line of sight (LOS) in parsecs. 
By combining equation A1 with equation A2, we obtain the $T_{b}$ for each pixel in the artificial bubbles with known $a_{\rm pc}$, $\Delta V$, and $n_{s}$, which are given by
 \begin{equation}
 \ T_{b}=2.615*10^{3}\frac{ n_{s}a_{\rm pc} }{\Delta V}.
 \label{equ:COColumndensity2}
 \end{equation}

We assume that the $^{13}$CO abundance relative to H$_2$ is 1.43 $\times$ 10$^{-6}$. 
The densities of $^{13}$CO affected by bubble expansion $n_0$ are $5 \times 10^{-5} cm^{-3}$ and $4 \times 10^{-4} cm^{-3}$ for Taurus and Perseus, respectively.
If all of the $^{13}$CO within the radius $R + \Delta R$ is distributed within the bubble, we can get $n_s$/$n_0$=1-${\xi}^{-3}$, where ${\xi}$=$(R+\Delta R)/R$. 

Figure 8 illustrates how $a_{\rm pc}$ is calculated. 
It shows a $^{13}\rm CO$ bubble sampled by five velocity channels in the left hemisphere as seen by an observer located to the far left. 
The starting and ending angles of the second velocity channel are $\theta_{i}$ and $\theta_{i+1}$, respectively. 
The angular dimension of each velocity channel is given by $\cos (\theta_{i+1}) = \cos (\theta_{i}) + \Delta V/V_{exp}$. 
Thereby, the angle of any LOS within the velocity channel can be determined. 
The physical depth $a_{\rm pc}$ of each LOS can be calculated from the angle of this LOS, bubble radius, and thickness. 
The detailed calculations are presented in Appendix A of Cazzolato $\&$ Pineault 2005.
By estimating the physical depth $a_{\rm pc}$ of each LOS for a given expansion velocity, and knowing $n_s$ of the bubble, we can obtain antenna temperature for each LOS using Equation A3. 
\begin{figure*}
\centering
%\begin{figure}[H]
%  % Requires \usepackage{graphicx}
\includegraphics[width=0.5\textwidth]{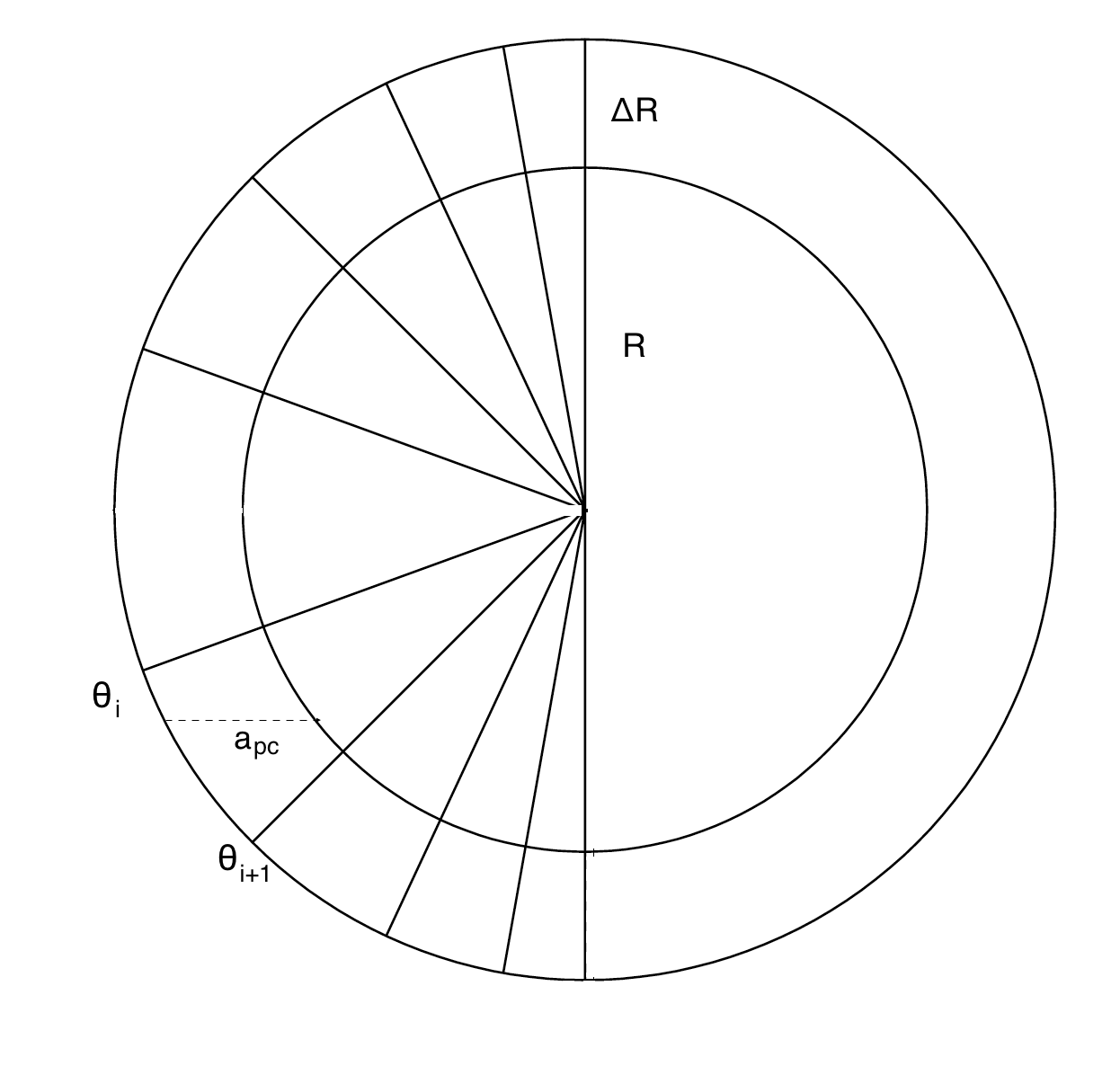}
\caption{$^{13}$CO bubble sampled by constant velocity channels with an expansion velocity of $V_{\rm exp}$. The radius and thickness of the bubble is R and $\Delta R$, respectively. An observer located far to the left would see a series of caps and rings of different depths. The dashed line shows the bubble physical depth at a certain line of sight. $\theta_{i}$ and $\theta_{i+1}$ are the starting angle and ending angle of the second velocity channel.}
% \label{modeling-brokenbubblePV}
 \end{figure*}

\section{Bubble detectability function derivation and fitting}
According to the definition, the experimental bubble detection result, y, is either 0 or 1 which corresponds to a Bernoulli distribution.
The probability mass function p of the detection results y is given by
\begin{eqnarray*}
p(y=1;\phi)&=&\phi,\\
p(y=0;\phi)&=&1-\phi,
\end{eqnarray*}
where $p(y=1;\phi)$ is the probability of detecting a bubble and $p(y=0;\phi)$ is the probability of not detecting a bubble.  
$p$ can be also expressed as \citep{NW72}
\begin{eqnarray*}
p(y;\phi)&=&\phi^{y} (1-\phi)^{1-y}\\
&=&\exp{(\log (\phi^{y}(1-\phi)^{1-y}) )}\\
&=&\exp{(y \log⁡ (\phi) +(1-y) \log⁡ (⁡1-\phi))}\\
&=&\exp{(\log (\dfrac{\phi}{1-\phi})y+\log (1-\phi))}.
\end{eqnarray*}
The exponential family is written as
\begin{eqnarray*}
f(y;\eta)=b(y)\exp{(\eta^{T} T(y)-a(\eta))},
\end{eqnarray*}
where $\eta$ is the natural parameter of the distribution; $T(y)$ is the sufficient statistic; and $a(\eta)$ is the log partition function.
When $\eta$ = $\log(\dfrac {\phi}{1-\phi})$, $a(\eta)=-\log(1-\phi)$, $T(y) = y$, and $b(y)=1$, we can derive the $p(y;\phi)$ from $f(y;\eta)$. This indicates that the probability mass function of the experimental bubble detection result is one of exponential family.
The experimental bubble parameters: $N$, $T_{a}$ and $V_{e}$ can be considered as independent to each other, so that we can combine them linearly after scaling. 
The scaling rule is as follows:
\begin{equation}
\mu_{scale} = \dfrac {\mu}{\mu_{max}}\\
\label{equ:er}
\end{equation}
where $\mu$ represents the bubble parameter, $\mu_{min}$ and $\mu_{max}$ refer to the minimum and the maximum value of parameter.

Thus, $\eta$ can be constructed as $\eta= \alpha^{T} X$. Here, $X$ denotes a three-dimensional vector containing $N$, $T_{a}$ and $V_{e}$. $\alpha$ denotes a four-dimensional coefficient vector containing an intercept $\alpha_{0}$. 
Our goal is to predict the expected value of the probability of detecting the bubble y with given $X$, 
which means we would like the detectability function $h(X)$ output to satisfy $h(X) = E[y|X]$.
In our formulation of the Bernoulli distribution as an exponential family distribution, we had $\eta = log(\dfrac {\phi}{1-\phi})$ which can be written as $\phi=\dfrac {1}{1+e^{-\eta}}$ and $E[y|X;\alpha]=\phi$.
So that the bubble detectability function $h(X)$ can be expressed as
\begin{eqnarray}
h(X)&=&E[y|X;\alpha] \\
&=&\phi \\
&=&\dfrac {1}{1+e^{-\alpha^{T} X}}.
\label{equ:er}
\end{eqnarray}
$h(X)$ is the logistic function. 
With given experimental bubble parameter and bubble detection result datasets, we can
 estimate the $\alpha^{T}$ for $h(X)$ based on the maximizing the likelihood function
$l(\theta)=\sum ^{rn}_{i=1} y^{(i)}log( h(X^{(i)}))+ (1-y^{(i)})log(1- h(X^{(i)})$, where rn is the row number of the data set.

\section{The kinetic energy probability density function fitting}
\label{sec:mlef}
We adopt the MLE to estimate the parameters of PDF of kinetic energy for both Taurus and Perseus using the truncated lognormal distribution.
The derivation for the MLE of the truncated lognormal distribution largely follows Zaninetti, L. 2017. 
Consider a sample  ${\mathcal X}=x_1, x_2, \dots , x_n$. 
The maximum and minimum value of sample can be expressed as $x_l$ and $x_u$, respectively, which are given by
\begin{equation}
{x_l}=\max(x_1, x_2, \dots, x_n), \qquad {x_u}=\min(x_1, x_2, \dots, x_n)
\quad  .
\label{eq:firstpar}
\end{equation}
The PDF can be expressed as
\begin{equation}
PDF (x;m,\sigma,x_l,x_u) = \frac
{
-\sqrt {2}{{\rm e}^{-\frac{1}{2}\,{\frac {1}{{\sigma}^{2}} \left( \ln  \left(
{\frac {x}{m}} \right )  \right ) ^{2}}}}
}
{
\sqrt {\pi}\sigma\, \left( {\rm erf} \left(a_{{3}}\right )-{\rm erf}
\left(a_{{8}}\right ) \right ) x
}
\quad,
\label{pdflognormaltruncatedcompact}
\end{equation}
where $m$ is the  scale parameter, $\sigma$ is the shape parameter, $a_{{3}}$ and $a_{{8}}$ are compact notations which are given by
\begin{equation}
a_3 =\frac{1}{2}\,{\frac {\sqrt {2} \left( \ln  \left( x_{{l}} \right ) -\ln
 \left( m \right )  \right ) }{\sigma}}
 \quad ,
 \nonumber
\end{equation}
\begin{equation}
a_8 = \frac{1}{2}\,{\frac {\sqrt {2} \left( \ln  \left( x_{{u}} \right ) -\ln
 \left( m \right )  \right ) }{\sigma}}
 \quad .
 \nonumber
\end{equation}
The CDF can be expressed as
\begin{equation}
CDF (x;m,\sigma,x_l,x_u)=
\frac
{
-{\rm erf} \left(\frac{1}{2}\,{\frac {\sqrt {2}}{\sigma}\ln  \left( {\frac {x
}{m}} \right ) }\right )+{\rm erf} \left(a_{{3}}\right )
}
{
{\rm erf} \left(a_{{3}}\right )-{\rm erf} \left(a_{{8}}\right )
}
\quad ,
\end{equation}

The MLE is obtained by maximizing
\begin{equation}
\Lambda = \sum_i^n \ln(PDF(x_i;m,\sigma,x_l,x_u)).
\end{equation}
The two derivatives $\frac{\partial \Lambda}{\partial m} =0$ and
$\frac{\partial \Lambda}{\partial \sigma}=0 $ generate two
nonlinear equations in $m$ and $\sigma$ which can be solved numerically, 
 \begin{eqnarray}
\frac{\partial \Lambda}{\partial m}=
   ( {\rm erf}   (\frac{1}{2}\,{\frac {\sqrt {2}   ( \ln    ( x_{{
l}}   ) -\ln    ( m   )    ) }{\sigma}}  )-
{\rm erf}   (\frac{1}{2}\,{\frac {\sqrt {2}   ( \ln    ( x_{{u}}
   ) -\ln    ( m   )    ) }{\sigma}}  )   )
  \nonumber \\
   ( n\sqrt {2}\sigma\,{{\rm e}^{-\frac{1}{2}\,{\frac {   ( \ln    (
x_{{l}}   ) -\ln    ( m   )    ) ^{2}}{{\sigma}^{2}}}}}
-n\sqrt {2}\sigma\,{{\rm e}^{-\frac{1}{2}\,{\frac {   ( \ln    ( x_{{u}}
   ) -\ln    ( m   )    ) ^{2}}{{\sigma}^{2}}}}}
\nonumber \\
-\sqrt
{\pi}   ( {\rm erf}   (\frac{1}{2}\,{\frac {\sqrt {2}   ( \ln
   ( x_{{l}}   ) -\ln    ( m   )    ) }{\sigma}}
  )
 \nonumber \\
  -{\rm erf}   (\frac{1}{2}\,{\frac {\sqrt {2}   ( \ln    ( x_{{
u}}   ) -\ln    ( m   )    ) }{\sigma}}  )   )
   ( n\ln    ( m   ) -\sum _{i=1}^{n}\ln    ( x_{{i}}
   )    )    ) =0
\quad ,
\end{eqnarray}
and
\begin{equation}
\frac{\partial \Lambda}{\partial \sigma}= \frac{N}{D} =0,
\end{equation}
where
\begin{eqnarray}
N = \ln  \left( x_{{u}} \right) \sqrt {2}{{\rm e}^{-\frac{1}{2}\,{\frac {   (
\ln    ( x_{{u}}   ) -\ln    ( m   )    ) ^{2}}{{
\sigma}^{2}}}}}n\sigma-\ln    ( x_{{l}}   ) \sqrt {2}{{\rm e}^{
-\frac{1}{2}\,{\frac {   ( \ln    ( x_{{l}}   )
-\ln    ( m
   )    ) ^{2}}{{\sigma}^{2}}}}}n\sigma
\nonumber \\
   +\sqrt {2}{{\rm e}^{-1/
2\,{\frac {   ( \ln    ( x_{{l}}   ) -\ln    ( m   )
   ) ^{2}}{{\sigma}^{2}}}}}\ln    ( m   ) n\sigma-\sqrt {2}
{{\rm e}^{-\frac{1}{2}\,{\frac {   ( \ln    ( x_{{u}}   ) -\ln
   ( m   )    ) ^{2}}{{\sigma}^{2}}}}}\ln    ( m
   ) n\sigma
\nonumber \\
   +n   ( \ln    ( m   )    ) ^{2}\sqrt {
\pi}{\rm erf}   (\frac{1}{2}\,{\frac {\sqrt {2}   ( \ln    ( x_{{u}}
   ) -\ln    ( m   )    ) }{\sigma}}  )
\nonumber  \\
-n{\sigma}^{
2}\sqrt {\pi}{\rm erf}   (\frac{1}{2}\,{\frac {\sqrt {2}   ( \ln
   ( x_{{u}}   ) -\ln    ( m   )    ) }{\sigma}}
  )
\nonumber \\
  -n   ( \ln    ( m   )    ) ^{2}\sqrt {\pi}
{\rm erf}   (\frac{1}{2}\,{\frac {\sqrt {2}   ( \ln    ( x_{{l}}
   ) -\ln    ( m   )    ) }{\sigma}}  )
\nonumber \\
+n{\sigma}^{
2}\sqrt {\pi}{\rm erf}   (\frac{1}{2}\,{\frac {\sqrt {2}   ( \ln
   ( x_{{l}}   ) -\ln    ( m   )    ) }{\sigma}}
  )
\nonumber \\
  +\sum _{i=1}^{n}\ln    ( x_{{i}}   )    ( \ln
   ( x_{{i}}   ) -2\,\ln    ( m   )    ) \sqrt {\pi}
{\rm erf}   (\frac{1}{2}\,{\frac {\sqrt {2}   ( \ln    ( x_{{u}}
   ) -\ln    ( m   )    ) }{\sigma}}  )
\nonumber \\
   -\sum _{i=1
}^{n}\ln    ( x_{{i}}   )    ( \ln    ( x_{{i}}   ) -
2\,\ln    ( m   )    ) \sqrt {\pi}{\rm erf}   (\frac{1}{2}\,{
\frac {\sqrt {2}   ( \ln    ( x_{{l}}   ) -\ln    ( m
   )    ) }{\sigma}}  )
\quad ,
\end{eqnarray}

\begin{eqnarray}
D=\sqrt {\pi}   \Bigg ( -{\rm erf}   
 \bigg (\frac{1}{2}\,{\frac {\sqrt {2}   ( \ln
   ( x_{{l}}   ) -\ln    ( m   )    ) }{\sigma}}
 \bigg )  
\nonumber \\ 
+{\rm erf}  \bigg  (\frac{1}{2}\,{\frac {\sqrt {2}   ( \ln    ( x_{{
u}}   ) -\ln    ( m   )    ) }{\sigma}} \bigg )  \Bigg )
{\sigma}^{3} \quad .
\end{eqnarray}


\begin{thebibliography}{}
\bibliographystyle{apj}
\bibliography{apj-jour,bibliography}

\bibitem[{Arce} {et~al.}(2011)]{Arce11}
Arce, H.~G., Borkin, M.~A., Goodman, A.~A., Pineda, J.~E., \&
Beaumont, C.~N.\ 2011, \apj, 742, 105

\bibitem[{Arce} {et~al.}(2010)]{Arce10}
Arce, H.~G., Borkin, M.~A., Goodman, A.~A., Pineda, J.~E., \&
Halle, M.~W.\ 2010, \apj, 715, 1170


\bibitem[{Arce} \& {Sargent}(2006)]{Arce06}
Arce, H.~G., \& Sargent, A.~I.\ 2006, \apj, 646, 1070
                                              

\bibitem[{Arce} \& {Goodman}(2002)]{Arce02}
Arce, H.~G., \& Goodman, A.~A.\ 2002, \apj, 575, 911


\bibitem[{Arce} \& {Goodman}(2001)]{Arce01}
Arce, H.~G., \& Goodman, A.~A.\ 2001, \apj, 554, 132
                                               
\bibitem[{Bally} {et al.}(2007)]{Bally07}
Bally, J., Reipurth, B., \& Davis, C. ~J. \ 2007, \planet, 951, 215

\bibitem[{Beaumont} {et al.}(2014)]{Beaumont14}
Beaumont, C.~N., Goodman, A.~A., Kendrew, S., et al.\ 2014, \apjs, 214, 3


\bibitem[{Cazzolato} \& {Pineault}(2005)]{Cazzolato05}
Cazzolato, F., \& Pineault, S.\ 2005, \aj, 129, 2731

\bibitem[{COMPLETE team} (2011)]{complete11}
COMPLETE team, \ 2011, \ https://hdl.handle.net/10904/10075, Harvard Dataverse, V2


\bibitem[{Churchwell} {et~al.}(2006)]{Churchwell06}
Churchwell, E., Povich, M.~S., Allen, D., et al. \ 2006, \apj, 649, 759


\bibitem[{Duarte-Cabral}  {et al.}(2012)]{Duarte-Cabral12}
Duarte-Cabral, A., Chrysostomou, A., Peretto, N. et al.\ 2012, \aap, 543, 140

%\bibitem[{D. J. Finney} (1941)]{finney1941}
%D. J. Finney. \ 1941 \aap, 543, 140
%Finney, D. J. , \ 1941, J. Roy. Statist. Soc. Ser. B, 7, 155


\bibitem[{Federrath, C.}(2015)]{Feder15}
Federrath, C., \ 2015, \mnras, 450, 4035

\bibitem[{Feddersen J. R.}{et al.}(2018)] {Feder18} 
Feddersen J. R., Arce, H\'ector G. , Shuo Kong, et al. \apj, 862, 121

\bibitem[{Frank} {et al.}(2014)]{Frank14}
Frank, A., Ray, T. P., Cabrit, S., et al.\ 2014, \planet, 914, 451

\bibitem[{Fuller} \& {Ladd}(2002)]{Fuller02}
Fuller, G.~A., \& Ladd, E.~F.\ 2002, \apj, 573, 699


\bibitem[{Fukui} {et al.}(1986)]{Fukui86}
Fukui, Y.,  Sugitani, K., Takaba, H., et al.\ 1986, \apj, 311, 85

\bibitem[{Goldsmith} {et~al.}(2008)]{Goldsmith08}
Goldsmith, P.~F., Heyer, M., Narayanan, G., et al.\ 2008, \apj, 680, 428

\bibitem[{James} {et al.}(2013)]{G.D.13}
G. James, D. Witten, T. Hastie, and R. Tibshirani, \ 2013, An introduction to statistical learning,
Vol. 112, Springer

\bibitem[{Heiles}(1979)]{Heiles79}
Heiles, C., \ 1979, \apj, 229, 533

\bibitem[{Hartmann} {et al.}(2001)]{Hartmann01}
Hartmann, L., Ballesteros-Paredes, J., \& Bergin, E. ~A.\ 2001, \apj, 562, 852

\bibitem[{Kroupa} {et al.}(2018)]{Pavel18}
Kroupa P., Je\v{r}\'{a}bkov\'{a} T., Dinnbier F., Beccari G., Yan Z., \ 2018, \aap, 612,
74

\bibitem[{Kwan} \& {Scoville}(1976)]{Kwan76}
Kwan, J., \& Scoville, N.\ 1976, \apjl, 210, l39

\bibitem[{Lada} \& {Harvey}(1981)]{Lada81}
Lada, C.~J. \& Harvey, P.~M.\ 1981, \apj, 245, 58


\bibitem[{Li} {et al.}(2015)]{Li15}
Li, H. X., Li, D., Xu, D., et al.\ 2015, \apjs, 219, 20

\bibitem[{Matzner}(2002)]{Matzner02}
Matzner, C.~D.\ 2002, \apj, 566, 302

\bibitem[{Mottram} {et al.}(2017)]{Mott17}
Mottram, J. C., van Dishoeck, E. F., Kristensen, L. E., et al. \ 2017, \aap, 600,
A99

\bibitem[{Nakamura} {et al.}(2011{\natexlab{a}})]{Nakamura11a}
Nakamura, F., Sugitani, K., Shimajiri, Y., et al.\ 2011, \apj, 737, 56

\bibitem[{Nakamura} {et~al.}(2011{\natexlab{b}})]{Nakamura11b}
Nakamura, F., Kamada, Y., Kamazaki, T., et~al.\ 2011, \apj, 726, 46

\bibitem[{Narayanan} {et~al.}(2008)]{Narayanan08}
Narayanan, G., Heyer, M.~H., Brunt, C., et al.\ 2008, \apjs, 177, 341

\bibitem[{Narayanan} {et~al.}(2012)]{Narayanan12}
Narayanan, G., Snell, R., \& Bemis, A. 2012, \mnras, 425, 2641

\bibitem[{Nelder}{Wedderburn}(1972)] {NW72}
Nelder J. A., Wedderburn R. W. M., 1972, Journal of the Royal Statistical
Society, Series A, General, 135, 370, DOI: https://doi.org/10.2307/2344614


\bibitem[{Norman} \& {Ikeuchi}(1989)]{Norman89}
Norman, C., \& Ikeuchi, S.\ 1989, \apj, 345, 372

\bibitem[{Offner} \& {Arce}(2015)]{Offner15}
Offner, S. S. R., \& Arce, H. G.\ 2015, \apj, 811, 146

\bibitem[{Plunkett} {et al.}(2013)]{Plunkett13}
Plunkett, A. ~L., Arce, H. ~G., Corder, S. ~A., et al.\ 2013, \apj, 774, 22

\bibitem[{Ridge}  {et al.}(2006a)]{Ridge06a}
Ridge, N. A., Di Francesco, J., Kirk, H., et al. \ 2006, \aj, 131, 2921

\bibitem[{T. Hastie} {et al.}(2001)]{The Elements of Statistical Learning} 
T. Hastie, R. Tibshirani, J. Friedman, The Elements of Statistical Learning,
Springer-Verlag, New York, \ 2001.

\bibitem[{Solomon} {et al.}(1981)]{Solomon81}
Solomon, P. ~M., Huguenin, G. ~R., \& Scoville, N. ~Z.\ 1981, \apj, 245, 19

\bibitem[{Snell} {et~al.}(1980)]{Snell80}
Snell, R.~L., Loren, R.~B., Plambeck, R.~L.\ 1980, \apj, 239, 17


\bibitem[{Zaninetti, L.}(2017)]{zani17}
Zaninetti, L. \ 2017, Advances in Astrophysics, 2, 197, DOI: 10.22606/adap.2017.23005

\bibitem[{Zinnecker} \& {Yorke}(2007)]{zinnecker07}
Zinnecker, H., \& Yorke, H.~W.\ 2007, \araa, 45, 481



\end{thebibliography}
\end{document}